\documentclass[aps,prf,onecolumn,10pt]{revtex4-1}
\usepackage{amsmath}
\usepackage{graphicx}
\usepackage{color}
\usepackage{hyperref}

\def \degre {$^\mathrm{o}$}
\def \mps {m~s$^{-1}$}

\usepackage{geometry}
\def \degre {$^\mathrm{o}$}
\def \mps {m~s$^{-1}$}

\begin{document}

\title{Wind-sustained viscous solitons}
\author{M. Aulnette}
\author{M. Rabaud}
\author{F. Moisy}
\email{frederic.moisy@u-psud.fr}
\affiliation{Laboratoire FAST, CNRS, Universit\'e Paris-Sud, Universit\'e Paris-Saclay, 91405 Orsay, France}

\date{\today}

\begin{abstract}
\vspace{0.4 cm}

When wind blows at the surface of a liquid of sufficiently high viscosity, a wave packet of small amplitude is first generated, which sporadically forms large-amplitude fluid bumps that rapidly propagate downstream. These nonlinear structures, first observed by Francis [Philos. Mag. {\bf 42}, 695 (1954)], have an almost vertical rear facing the wind and a weak slope at the front. We call them  viscous solitons.  We investigate their dynamics in a wind-tunnel experiment using silicon oil of kinematic viscosity 1000~mm$^2$~s$^{-1}$ by means of laser sheet profilometry and particle image velocimetry. We give evidence of their subcritical nature: they are emitted in a region of large shear stress but, once formed, they are sustained by the wind and propagate in a region of lower stress. Their propagation velocity is given by the balance between aerodynamic drag in the air and viscous drag in the liquid.  The stable soliton branch of the subcritical bifurcation diagram is reconstructed from the measured soliton amplitude at various wind velocities and distances along the channel. At large wind velocity, the emission frequency of solitons increases, resulting in a long-range sheltering of downstream mature solitons by newly formed upstream solitons,  which limits their course.

\end{abstract}

\maketitle

\section{Introduction} \label{sec:intro}

When wind blows over water, small-amplitude disordered wrinkles elongated in the wind direction first appear, which, for a wind velocity above a critical value, turn into growing waves propagating predominantly in the direction of the wind~\cite{Kahma_1988,caulliez1999three,lin2008direct,Paquier_2015,zavadsky2017two,Perrard2019,hwang2019wind}.
The amplitude of these waves slowly increases with wind velocity and with downstream distance, leading to a gradual nonlinear evolution of their spectral content: Their typical wavelength increases, starting from about twice the capillary length at the onset, while their typical frequency decreases~\cite{dias1999nonlinear,caulliez1999three,veron2001experiments,zavadsky2017investigation}.

Increasing the viscosity of the liquid up to 100 times the water viscosity does not change this picture: Above a critical wind velocity, which slightly increases with the liquid viscosity, the initially sinusoidal wave train slowly evolves towards a more complex pattern at large downstream distance \cite{Paquier_2016}. Importantly, this nonlinear evolution remains moderate, in the sense that the spatial growth rate of the initial wave train remains much smaller than the typical wavelength.

\begin{figure}[t]
	\begin{center}
\includegraphics[width=8 cm]{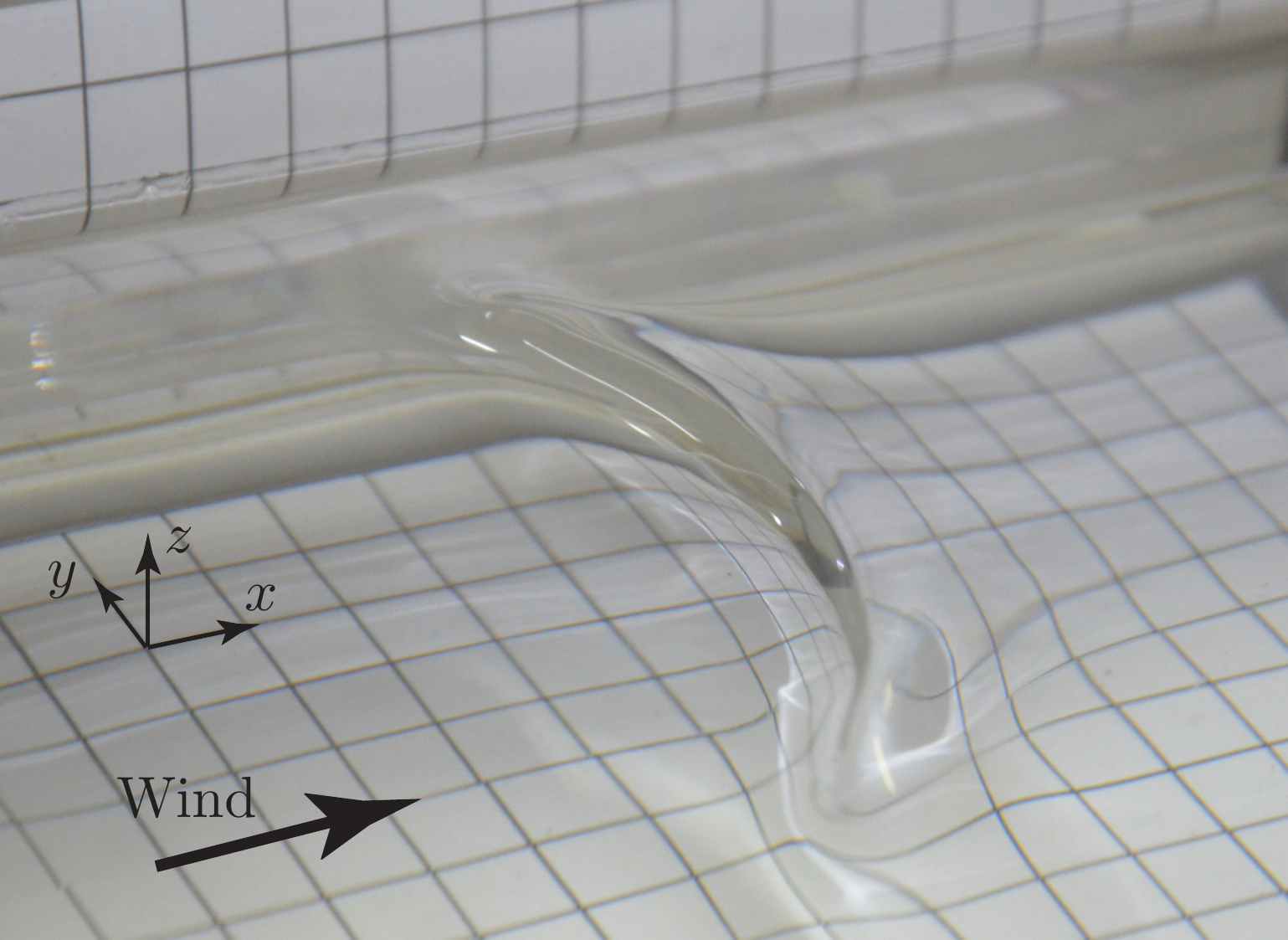}
\caption{Viscous soliton propagating from left to right, generated by a wind of velocity $U_a = 9.63$~\mps \ on a silicon oil bath of viscosity $\nu_\ell = 1000$ mm$^2$~s$^{-1}$ and depth $h = 35$~mm. The size of the grid pattern is 1.3 cm.}
\label{fig:PhotoSoliton}
\end{center}
\end{figure}

However, for a liquid of sufficiently high viscosity, this picture dramatically changes:  the initial wave train becomes strongly unstable even very close to the wind velocity threshold, and it rapidly evolves over a distance of the order of one wavelength into a large fluid bump pushed by the wind.  Such a  viscous soliton is illustrated in Fig.~\ref{fig:PhotoSoliton} for a liquid kinematic viscosity of 1000~mm$^2$~s$^{-1}$ and a wind velocity of $9.6$~\mps \ (see below for experimental details).  Viscous solitons are typically $2-4$ mm high, $10-20$ mm wide in the streamwise direction, with an almost vertical rear facing the wind and a weak slope at the front. Their spanwise extent is about 8~cm close to the critical wind velocity and increases up to the channel width (30~cm) at larger wind velocity, with a slightly curved crest line.

Viscous solitons provide a striking example of out-of-equilibrium coherent structures, resulting from the balance between an external forcing and dissipation~\cite{christov1995dissipative,Knobloch2008,Knobloch2015}. Note that although the viscous solitons considered here and the classical inviscid solitons in weakly dispersive shallow water waves are both self-preserving nonlinear objects, they are of a very different nature: Viscous solitons are strongly dissipative objects, which would rapidly decay without the continuous supply of energy by the wind.

Viscous solitons forced by the wind were first reported by Francis, in a series of experiments using  oils and syrups of kinematic viscosity $\nu_\ell$ from 250 to 58~000~mm$^2$~s$^{-1}$ \cite{Francis_1954,francis1956lxix}. Reproductions of his photographs can be found in Ref.~\cite{Miles1959generation} for the linear wave regime, and in Ref.~\cite{chandrasekhar} for the nonlinear soliton regime. Surprisingly, since the early work of Francis, these viscous solitons were not investigated further until the recent experiments of Paquier~{\it et al.} \cite{Paquier_2016,Paquier_PhD_2016}. In these experiments, liquids in a range of viscosity $\nu_\ell \simeq 1-560$~mm$^2$~s$^{-1}$ were used, covering the transition between the classical low-viscosity wind waves and the high-viscosity solitons. Both Francis \cite{Francis_1954,francis1956lxix} and Paquier~{\it et al.} \cite{Paquier_2016,Paquier_PhD_2016} note that, once it appears, the initial wave packet is systematically unstable and forms solitons, and that the critical wind velocity for their generation, of the order of $9-11$~\mps, shows almost no dependence on liquid viscosity. Another remarkable feature of viscous solitons is their finite amplitude even very close to the critical wind velocity, suggesting that they arise from a subcritical instability.

The conditions under which the interface between a viscous liquid and a low viscosity gas flow becomes unstable have been the subject of extensive literature (see Ref.~\cite{boomkamp1996classification} for a review). Stability analyses are usually based on either the thin-film approximation or  the deep-water approximation. The instability that leads to viscous solitons fall in the second category, to which we restrict our attention in the following. A central issue in this problem is the relevance of the Kelvin-Helmholtz instability mechanism: While it is well established that this mechanism does not describe the wave generation in the low-viscosity case, including in the air-water configuration, it is is usually considered as the relevant mechanism in the large-viscosity case~\cite{taylor1940generation, Miles1959generation, chandrasekhar, Hogan1985, barnea1993kelvin, kim2011viscous}. The reason for this somewhat paradoxical result is that although the liquid viscosity affects the growth rate of the instability, it has no effect on the critical wind velocity or on the most unstable wavelength, which remain governed by the inviscid Kelvin-Helmholtz predictions.  Such a velocity threshold independent of the liquid viscosity is indeed consistent with the observations of Francis~\cite{Francis_1954,francis1956lxix} and Paquier~{\it et al.} \cite{Paquier_2016,Paquier_PhD_2016}. Accordingly, viscous solitons can be seen as the nonlinear saturated state that results from the Kelvin-Helmholtz instability on a liquid of sufficient viscosity. This conclusion also holds in the closely related problem of slug formation in two-phase pipe flows, a configuration of primary interest in the oil industry~\cite{andritsos1989effect}.

The aim of this paper is to characterize experimentally the viscous soliton regime of Francis~\cite{Francis_1954}. Based on the same experimental setup as in Ref.~\cite{Paquier_2015}, experiments were carried out in this regime, using silicon oil of viscosity $1000$~mm$^2$~s$^{-1}$ and various liquid depths.  Our experiments confirm the subcritical nature of the instability that leads to the generation of solitons: We show that they are emitted in a region of large wind stress (thin air boundary layer), but, once formed, they can propagate in regions of lower wind stress (thicker boundary layer). We characterize in detail their properties (shape, amplitude, and velocity) as a function of the air velocity and liquid depth, and show that their propagation results from a balance between the aerodynamic drag in the air and the viscous drag in the liquid.  We finally discuss the conditions under which viscous solitons may form when blowing over a liquid surface.

\section{Experimental setup and flow characterization}
\label{sec:exp_setup}

\subsection{Wind-tunnel and liquid tank}

\begin{figure}[t]
\begin{center}
\includegraphics[width=13cm]{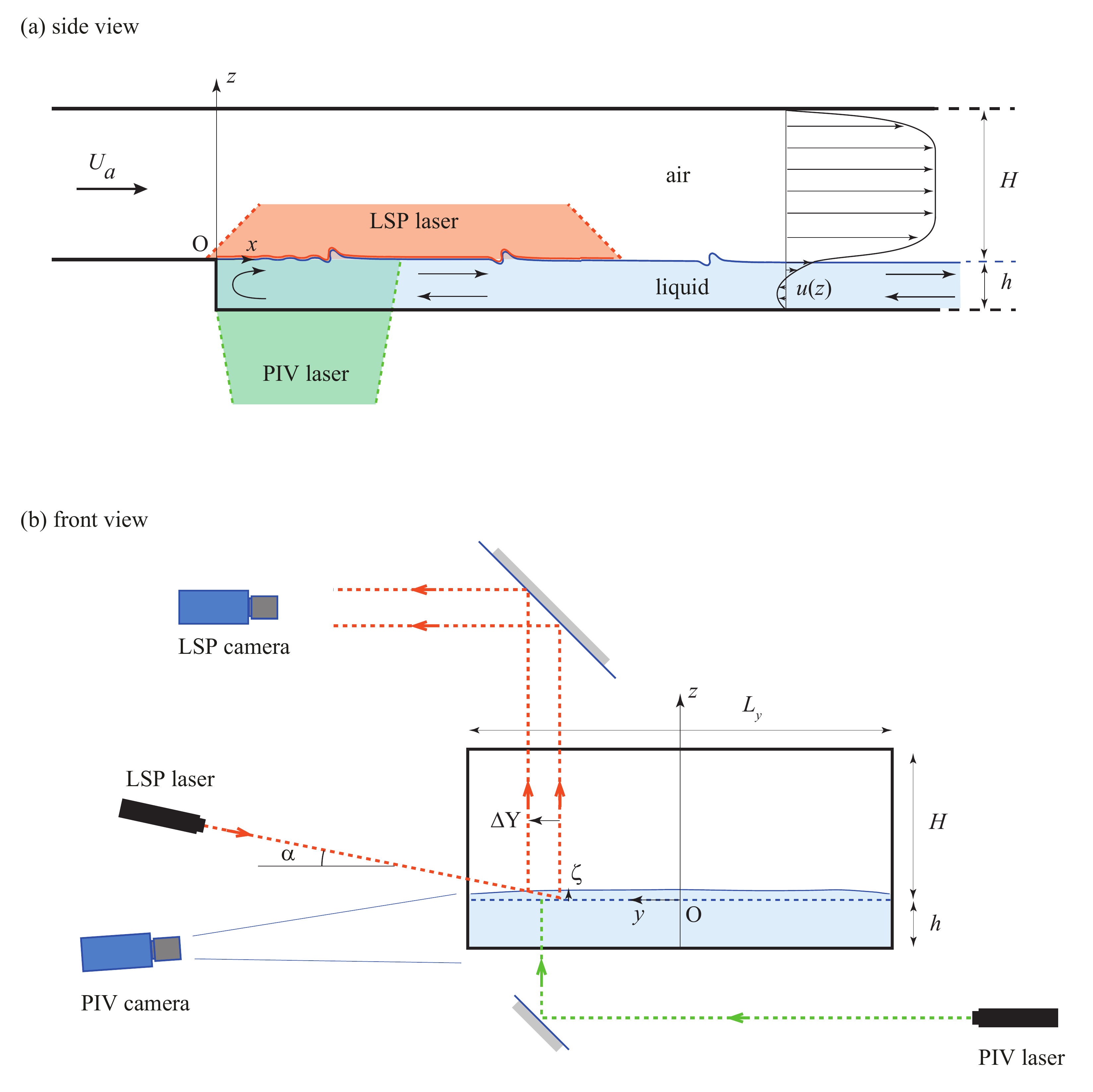}
\caption{Sketch of the wind tunnel and of the liquid tank: (a) side view and (b) front view, looking downstream. Measurements are performed by laser sheet profilometry and particle image velocimetry. The wave amplitude is exaggerated for visibility.}
\label{fig:Set-up}
\end{center}
\end{figure}

The experimental setup is composed of a liquid-filled rectangular tank of length $L_x = 1.5$~m, width $L_y= 296$~mm, located at the bottom of a wind tunnel (Fig.~\ref{fig:Set-up}). The depth  $h$ of the liquid tank is varied in the range $20-50$~mm using immersed Plexiglas plates at the bottom. The  wind tunnel height is $H=$ 105 mm, and its width is equal to the width of the liquid tank.

The working fluid is silicon oil (Bluesil Fluid 47 V1000), of density $\rho_\ell = 970$ kg~m$^{-3}$, kinematic viscosity $\nu_\ell =$ 1000 mm$^2$~s$^{-1}$,  and surface tension $\gamma =$ 21.1~mN~m$^{-1}$ at $25$\degre. Using silicon oil, a non-polar liquid with low surface tension, reduces surface pollution, which is unavoidable with aqueous solutions.  The air flow, generated by a wind turbine located upstream, can be adjusted in the range $U_a = 1-15$~m~s$^{-1}$. The wind velocity here is limited to 11 m~s$^{-1}$; otherwise, solitons reach the end of the tank, resulting in a loss of oil.   More details on the setup and properties of the air boundary layer can be found in Ref. \cite{Paquier_PhD_2016}.

Two measurement methods are used: laser sheet profilometry (LSP) (Fig.~\ref{LSP}) and particle image velocimetry (PIV) (Fig.~\ref{PIV}). Both measurements are performed along $x$, at a distance of $45$~mm from the left side wall of the channel.  For LSP, the oil is made diffusive by adding TiO$_2$ powder (a white pigment) at a concentration of 250 mg/l. For PIV, the oil is seeded with coated glass beads, 10 $\mu$m in diameter, at a concentration of 25 mg/l. Because of these different seedings, the two methods cannot be used simultaneously.

\subsection{LSP measurements}

\begin{figure}[t]
\begin{center}
\includegraphics[width=14 cm]{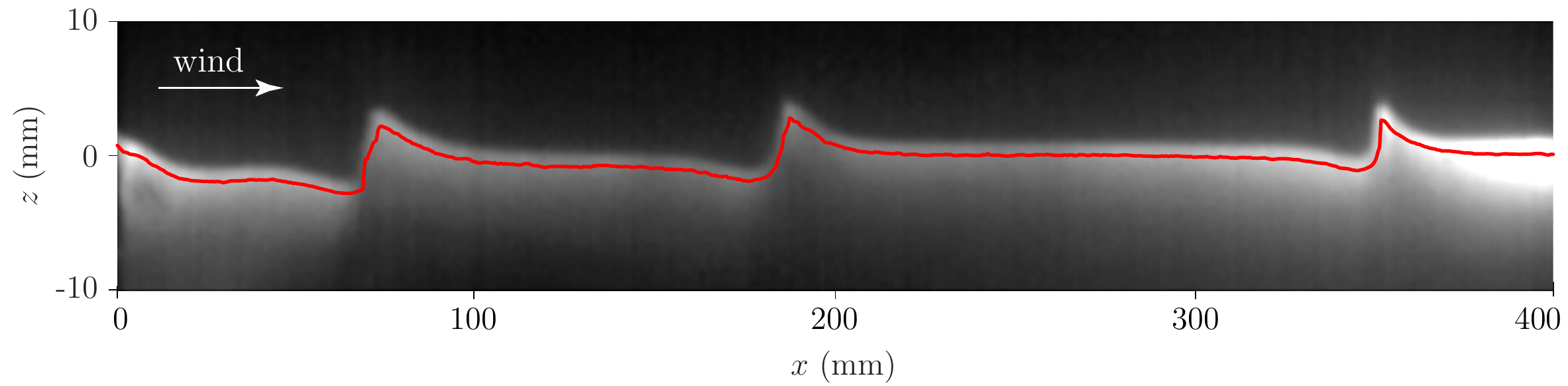}
\caption{Free surface measurement using LSP showing three viscous solitons propagating downstream, for $U_a = 9.95$~\mps \ and $h = 35$~mm. The grayscale image shows the raw image of the tilted laser sheet penetrating the oil;  the solid red line is the reconstructed interface $\zeta(x)$. Note that the vertical scale is magnified for visibility.}
\label{LSP}
\end{center}
\end{figure}

For LSP measurements, the liquid is illuminated by a laser sheet making a small incidence angle $\alpha$ with the horizontal, and the intersection between the laser sheet and the surface is imaged from above with a 1280$\times$1024 camera (PCO 1200hs mounted with a 85 mm Nikon lens) working at 20 frames/s. For a surface deformation $\zeta(x,t)$ invariant along $y$, this intersection is shifted horizontally by a distance $\Delta Y(x,t)  = \zeta(x,t) / \tan \alpha$ [see Fig.~\ref{fig:Set-up}(b)]. The angle $\alpha$ is chosen small to increase the vertical resolution, but such that the typical deviation $\Delta Y$ remains smaller than the transverse curvature of the solitons (their spanwise extent is of the order of 8~cm at small wind velocity, as shown in Fig.~\ref{fig:PhotoSoliton}, and up to the channel width at larger wind velocity). In practice, we chose $\alpha = 16$\degre, yielding $\Delta Y$ of the order of 1~cm for a 3-mm soliton.

The surface elevation $\zeta(x,t)$ is determined by a standard edge detection method: The maximum light intensity at each $x$ is determined with subpixel accuracy using a 11-point parabolic fit of the light intensity, yielding a vertical resolution of 0.1~mm. A typical LSP image is shown in Fig.~\ref{LSP}. Because of the pressure drop in air along the tunnel, the liquid surface is not perfectly horizontal but shows a marked lowering at $x=0$ of typically 1~mm, followed by a shallow positive slope along the channel. This lowering is more pronounced at small liquid depth, which limits the measurements to $h \geq 20$~mm. In the following, this mean surface profile is substracted, and we define the wave profile as $\xi(x,t)=\zeta(x,t)-\langle \zeta(x,t) \rangle_t$, with $\langle \cdot \rangle_t$ a time average.  Note that for strong forcing the upstream side of the soliton shows an overhanging profile, so $\zeta(x)$ is no longer monovalued. Because of the observation from above, only the upper part of the overhang can be detected, and no reliable measurement can be performed where the slope is nearly vertical.

From the LSP measurements, solitons are identified using a two-step tracking algorithm: (i)~In the identification step, for each time $t$ we detect the locations of the maxima of the surface height $\xi(x,t)$, with one maximum per interval in which $\xi(x,t)$ exceeds the threshold of 0.5~mm. A minimum distance of $8$~mm is taken between two consecutive maxima. (ii)~In the pairing step, the solitons detected in consecutive frames are paired using a standard nearest neighbor scheme. Using this algorithm, we build a set of $N$ solitons for each wind velocity, with $N\simeq 30$ at small $U_a$, up to $N\simeq 100$ at large $U_a$. For each soliton, we record its location $X_s(t)$, velocity $V_s(t) = d X_s / dt$, and amplitude $A(t) = \max \xi$.

\subsection{PIV measurements}

\begin{figure}[t]
\begin{center}
\includegraphics[width=11 cm]{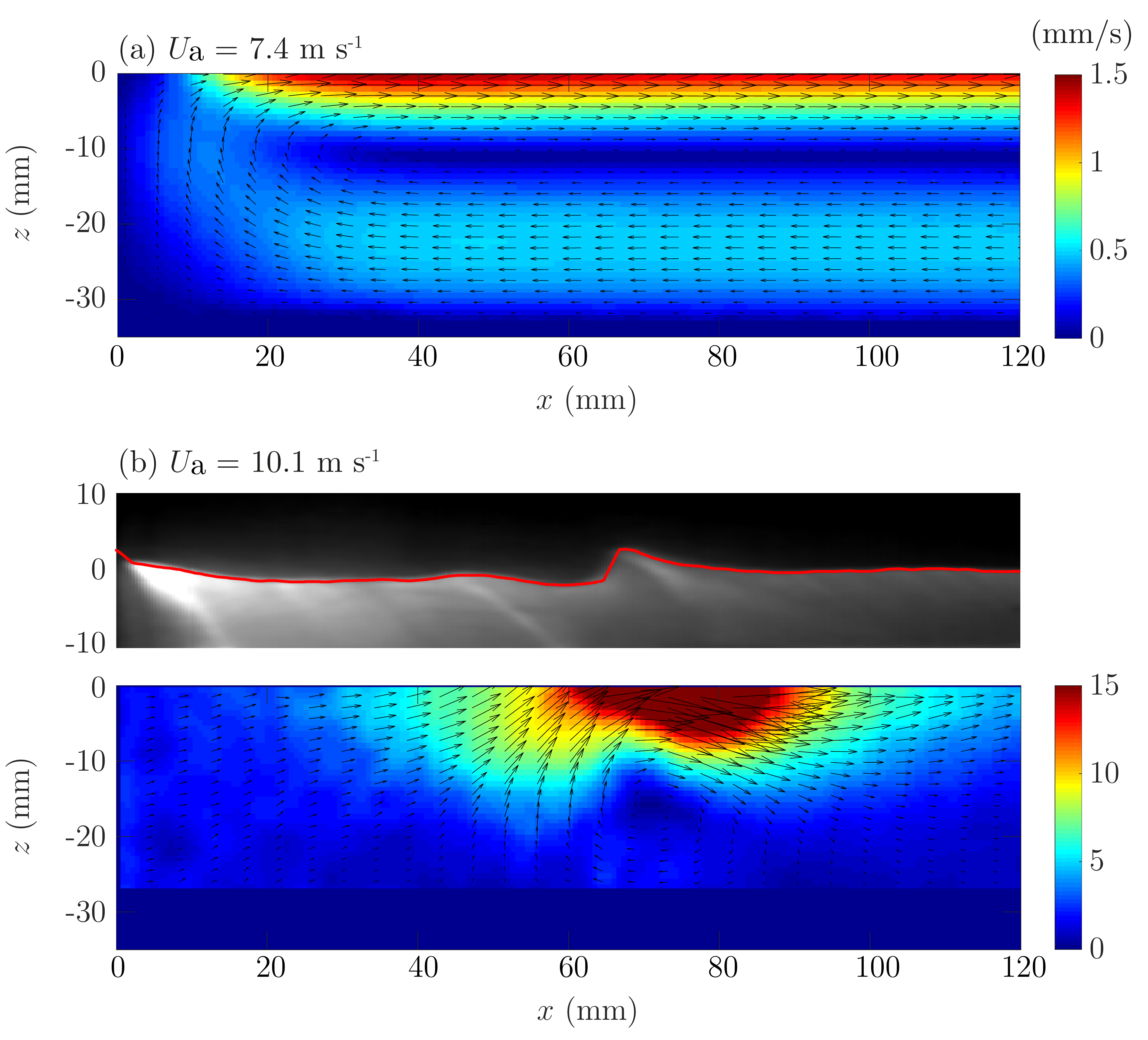}
\caption{Flow in a vertical plane measured by PIV. (a), below the onset of solitons, $U_a=7.4$~\mps, and (b) showing the formation of a soliton, $U_a = 10.1$~m~s$^{-1}$.  Color codes the norm of the velocity. Because of the camera incidence, the 5 mm at the bottom of the tank are not resolved. In (b) a soliton measured by LSP is also shown at the same location as in the PIV field (PIV and LSP measurements are not simultaneous).}
\label{PIV}
\end{center}
\end{figure}

Particle image velocimetry measurements are performed in a vertical plane,  right after the edge of the tank, for $x=0-120$~mm, where solitons are generated (see Fig.~\ref{fig:Set-up}). The flow is illuminated from below, to avoid deformation from the surface, using a vertical laser sheet from a double-pulsed Nd:YAG laser (25 mJ/pulse).  The repetition rate of the image pairs is 10~Hz, and the time interval between two images of a pair is set between 10 and 100~ms.   Images are taken from the side using a double-frame camera (PCO 2000 mounted with a 35 mm Nikon lens), making a small angle of 5$^\mathrm{o}$ in order to better resolve the flow close to the free surface; however, because of the curvature of the crest of the solitons, the flow cannot be resolved inside the soliton bump, and measurements are restricted to $z<0$. The velocity fields are computed using standard PIV algorithm, with interrogation windows of 16~pixels, yielding a spatial resolution of 0.7~mm.

Two PIV fields are shown in Fig.~\ref{PIV}, one below and one above the onset of solitons. The velocity field below the onset [Fig.~\ref{PIV}(a)] shows a surface drift velocity induced by the wind shear stress, of order of $1.5$~mm~s$^{-1}$, and a backflow underneath.  At larger wind velocity [Fig.~\ref{PIV}(b)], in addition to this mean recirculation flow, we observe the emission of solitons, characterized by a vortex traveling below the surface bump (PIV and LSP measurements are not performed simultaneously, but the snapshots shown here correspond to a similar event). The fluid velocity below a soliton is of the order of its propagation velocity, here about 40~mm~s$^{-1}$, which is at least 20 times faster than the mean recirculation flow.

\subsection{Spatial decay of the wind shear stress}

An important parameter for the generation and propagation of solitons in this experiment is the rate of decrease of the wind shear stress $\tau(x)$ at the surface. This spatial decay is related to the increase of the  turbulent boundary layer thickness $\delta(x)$~\cite{Schlichtling}.

\begin{figure}[t]
	\begin{center}
\includegraphics[width=12cm]{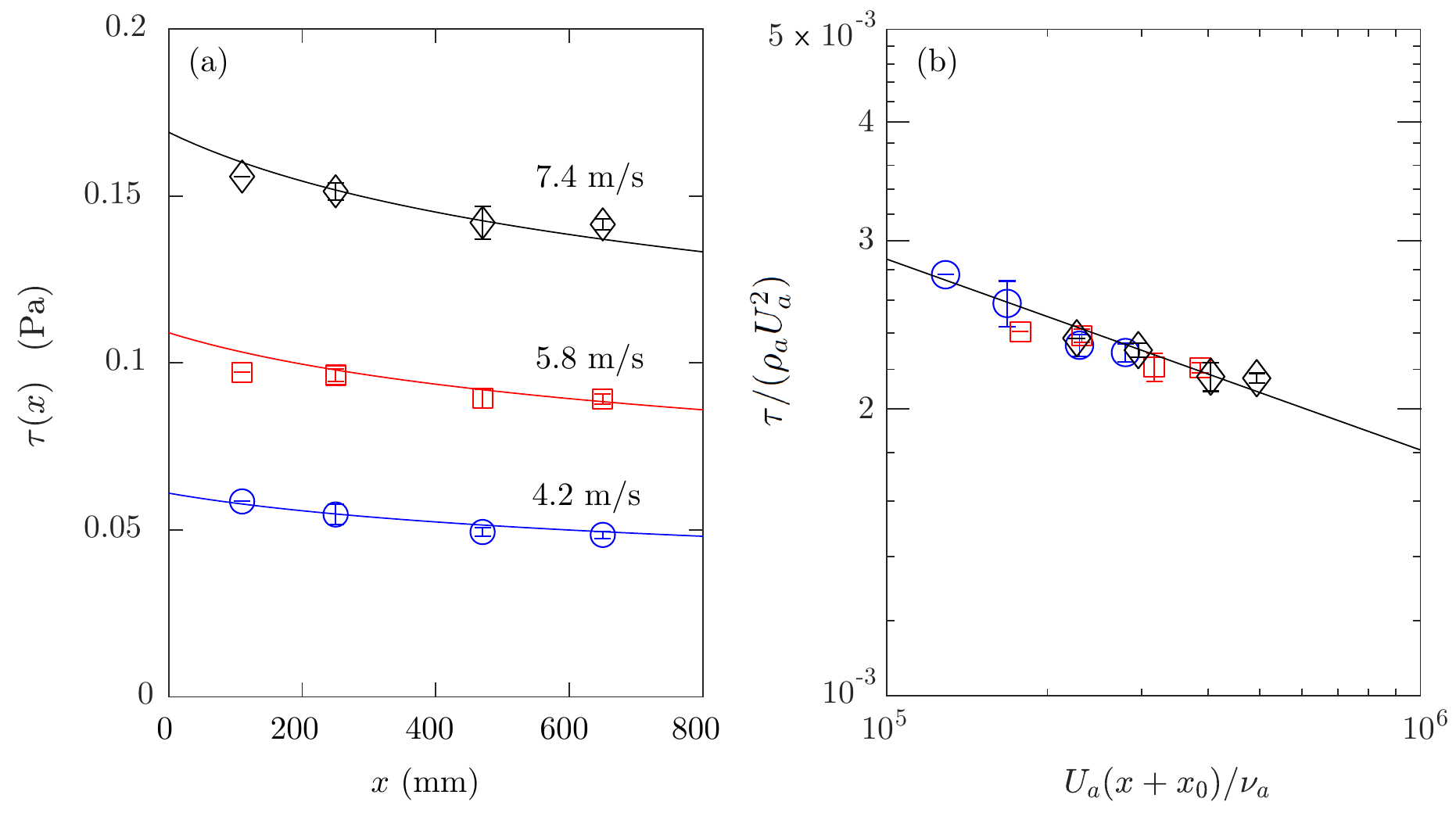}
\caption{(a) Shear stress $\tau$ as a function of the distance $x$ for three wind velocities $U_a$, obtained from the drift velocity measurement using Eq.~(\ref{eq:us}). (b) Normalized shear stress (skin friction coefficient) as a function of the Reynolds number based on the total streamwise distance $x+x_0$, where $x_0$ is the length of the flat plate before the edge of the tank. In both figures the lines show the empirical fit [Eq.~(\ref{eq:tauvsx})].}
\label{ustar}
\end{center}
\end{figure}

Below the onset of solitons, the local shear stress $\tau(x)$ can be inferred from the local drift velocity $U_s(x)$ using stress continuity at the interface:  Assuming an applied shear stress slowly varying in $x$, the Stokes flow of a viscous liquid in a closed container of thickness $h$ is parabolic,
\begin{equation}
u(x,z) \simeq U_s(x) \left(1+\frac{z}{h} \right) \left(1 + 3\frac{z}{h} \right),
\label{eq:par}
\end{equation}
with $-h \leq z \leq 0$, where $U_s(x)$ is the surface drift velocity
\begin{equation}
U_s(x) = \frac{\tau(x) h}{4 \rho_\ell \nu_\ell}.
\label{eq:us}
\end{equation}
The recirculation flow in Fig.~\ref{PIV}(a) is consistent with the parabolic profile (\ref{eq:par}), at least for $x$ not too close to the edge of the tank (typically for $x > h$) and $y$ not too close to the side walls. The local shear stress deduced from this measurement, plotted in Fig.~\ref{ustar}(a), clearly shows a decrease in $x$. It can be described in terms of the skin friction coefficient, $\tau/(\rho_a U_a^2)$, as a function of the Reynolds number based on the total streamwise distance $x+x_0$. Here $x_0 = 350$~mm is the length of the flat plate between the end of the wind-tunnel convergent and the edge of the liquid tank at $x=0$. The data show excellent agreement with the classical empirical fit for a developing turbulent boundary layer [Fig.~\ref{ustar}(b)]
\begin{equation}
\frac{\tau(x)}{\rho_a U_a^2} \simeq C \left( \frac{U_a (x+x_0)}{\nu_a} \right)^{-0.2}.
\label{eq:tauvsx}
\end{equation}
The value of the coefficient $C = 0.029$ is identical to that obtained for a conventional no-slip boundary layer over a rigid wall~\cite{Schlichtling}. This is because of the very small drift velocity $U_s$ (less than 2~mm~s$^{-1}$), which is at least three orders of magnitude smaller than $U_a$. This law applies only below the onset of waves and solitons, when the influence of the wave roughness on the turbulence can be neglected.

\section{The life cycle of viscous solitons} 
\label{sec:lcs}

\begin{figure}
\begin{center}
\includegraphics[width=14cm]{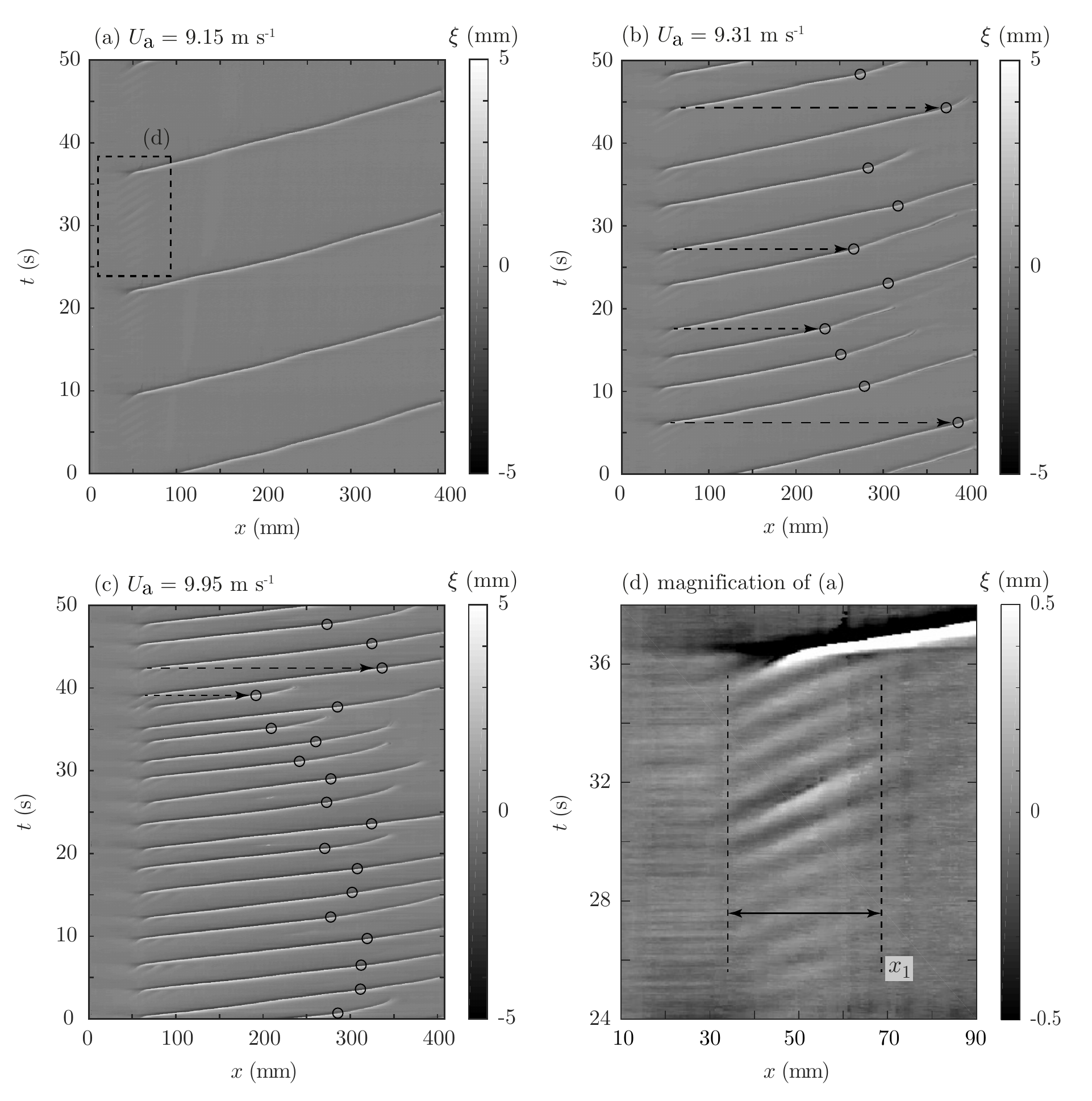}
\caption{Spatio-temporal diagrams of the surface elevation $\xi(x,t)$ during 50 s for a liquid depth $h=35$~mm at three wind velocities, showing the emission, propagation, and decay of viscous solitons.  The dashed box in (a), magnified in (d), shows the interval of $x$ where the initial wave packet is present. The horizontal dashed arrows and the circles highlight the influence of new upstream solitons on mature downstream solitons: strong sheltering at moderate wind velocity, leading to a marked slowdown sometimes followed by a decay in (b), and weak sheltering at larger wind velocity in (c).}
\label{spatiotemp}
\end{center}
\end{figure}

We now turn to the generation of viscous solitons, and provide here a general overview of their life cycle. It can be divided in three phases: generation, propagation and decay.  These three phases are visible in the spatio-temporal diagrams of the surface elevation $\xi(x,t)$ shown in Fig.~\ref{spatiotemp}, for a liquid depth $h=35$~mm and 3 values of the wind velocity $U_a$. For this liquid depth, the critical wind velocity is $U_{ac} = 9.15$~\mps.

We first note that for $U_a$ smaller than the critical velocity $U_{ac}$, the surface remains essentially stationary at all $x$: The wrinkles (random wakes produced by the pressure fluctuations in the turbulent boundary layer \cite{Perrard2019}) are of amplitude less than 2~$\mu$m for such a large viscosity, which is well below the resolution of the LSP measurements. 

\begin{figure}
\begin{center}
\includegraphics[width=8cm]{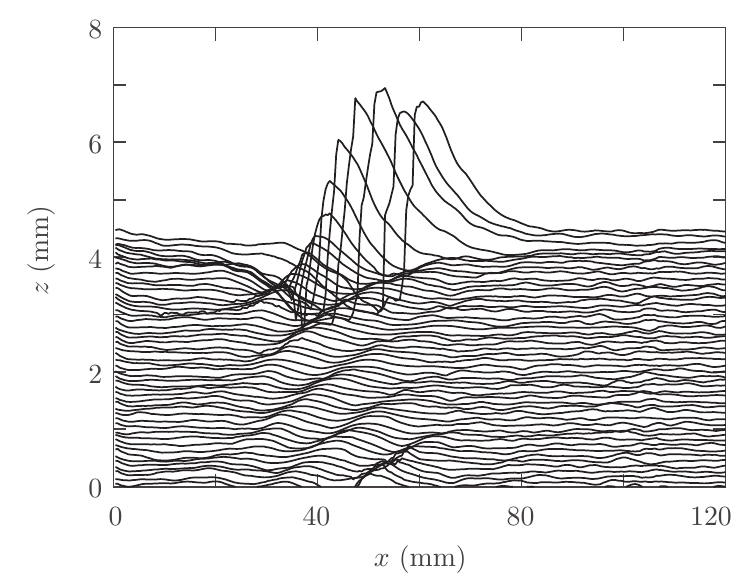}
\caption{Stacked plots of the surface elevation $\xi(x)$ for $U_a = 9.15$~\mps \ taken every $\Delta t = 0.1$~s, showing the propagation and growth of the initial wave packet and the generation of a soliton. The profiles are vertically shifted by $0.1$~mm at each time step.}
\label{stacked}
\end{center}
\end{figure}

As the wind velocity is increased above $U_{ac}$, a wave packet composed of a few propagative waves of small amplitude is formed in a narrow interval of $x$ [see the magnification in Fig.~\ref{spatiotemp}(d)]. This interval is approximately $30-70$~mm close to the onset, and increases up to $30-100$~mm at larger velocity. The wave packet remains spatially confined in this interval, but it is always unstable in time: Its amplitude grows, at a rate that increases with the wind velocity, until one crest abruptly steepens and forms a large amplitude soliton. This process is better visualized in the stacked plot shown in Fig.~\ref{stacked}. Once formed, the soliton accelerates, leaves the initial wave packet, and propagates along the channel for a certain distance.  After the formation of a soliton, the wave packet disappears for some time and then reforms and the process repeats.

\begin{figure}[t]
\begin{center}
\includegraphics[width=14 cm]{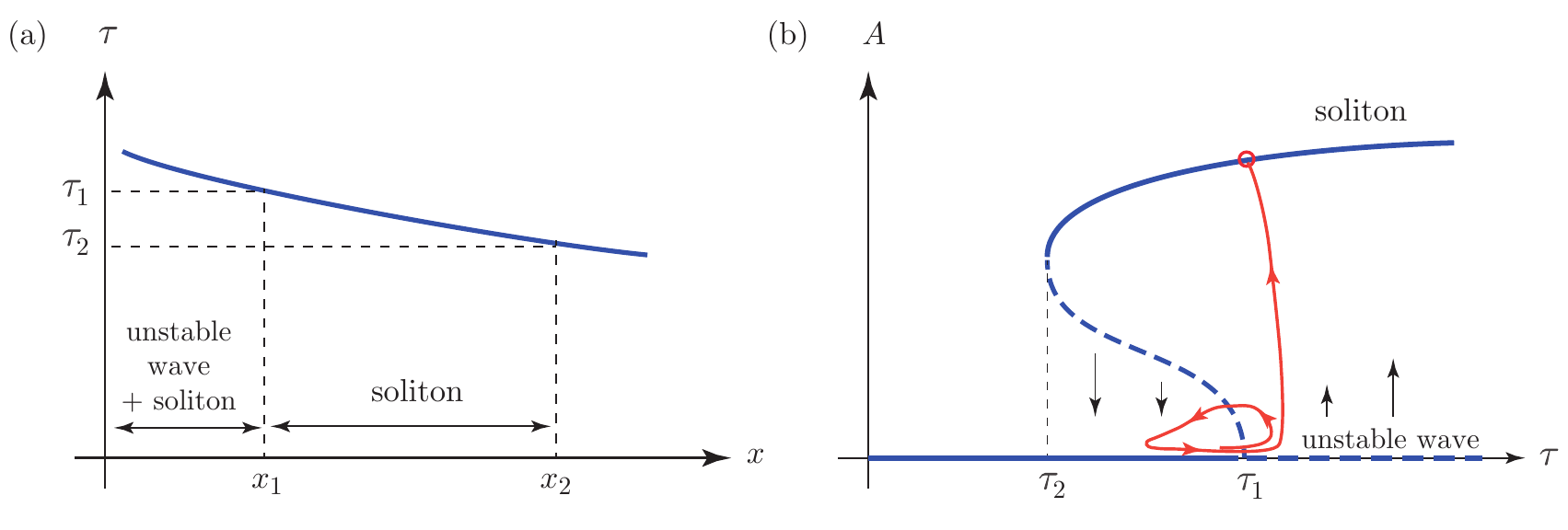}
\caption{Sketch illustrating the subcritical formation of viscous solitons. (a) Decrease of the shear stress along the channel, $\tau \sim (x+x_0)^{-0.2}$ [Eq.~(\ref{eq:tauvsx})], and the two thresholds $\tau_1 > \tau_2$ defining the two distances $x_1 < x_2$. (b) Bifurcation diagram, showing the primary wave branch, unstable for $\tau > \tau_1$, and the soliton branch, stable for $\tau > \tau_2$ (stable branches in solid lines, unstable branches in dashed lines). The vertical black arrows illustrate the linear growth rate $\sigma_i$. The red trajectory illustrates the evolution of the system when governed by a time-varying shear stress $\tau$ close to the onset $\tau_1$ (see Sec.~IV).}
\label{fig:sketch}
\end{center}
\end{figure}

These observations suggest a description of the generation and propagation of solitons in terms of a subcritical instability governed by the local shear stress $\tau(x)$. Figure~\ref{fig:sketch} illustrates this scenario. For a given wind velocity $U_a$, the fact that the initial wave packet is found only up to a certain distance $x_1$ indicates that it is sustained only for a sufficient shear stress $\tau_1 = \tau(x_1)$ [see Fig.~\ref{fig:sketch}(a)]. Beyond $x_1$, the shear stress decreases below $\tau_1$ and the wave packet disappears. However, once formed, solitons can cross the boundary $x_1$ and propagate in the region of lower shear stress $\tau(x)<\tau_1$.  Solitons in the subcritical region are sustained until the local shear stress decreases below a secondary threshold $\tau_2 < \tau_1$, corresponding to the distance $x_2$ of decay of solitons.

\begin{figure}
\begin{center}
\includegraphics[width=8cm]{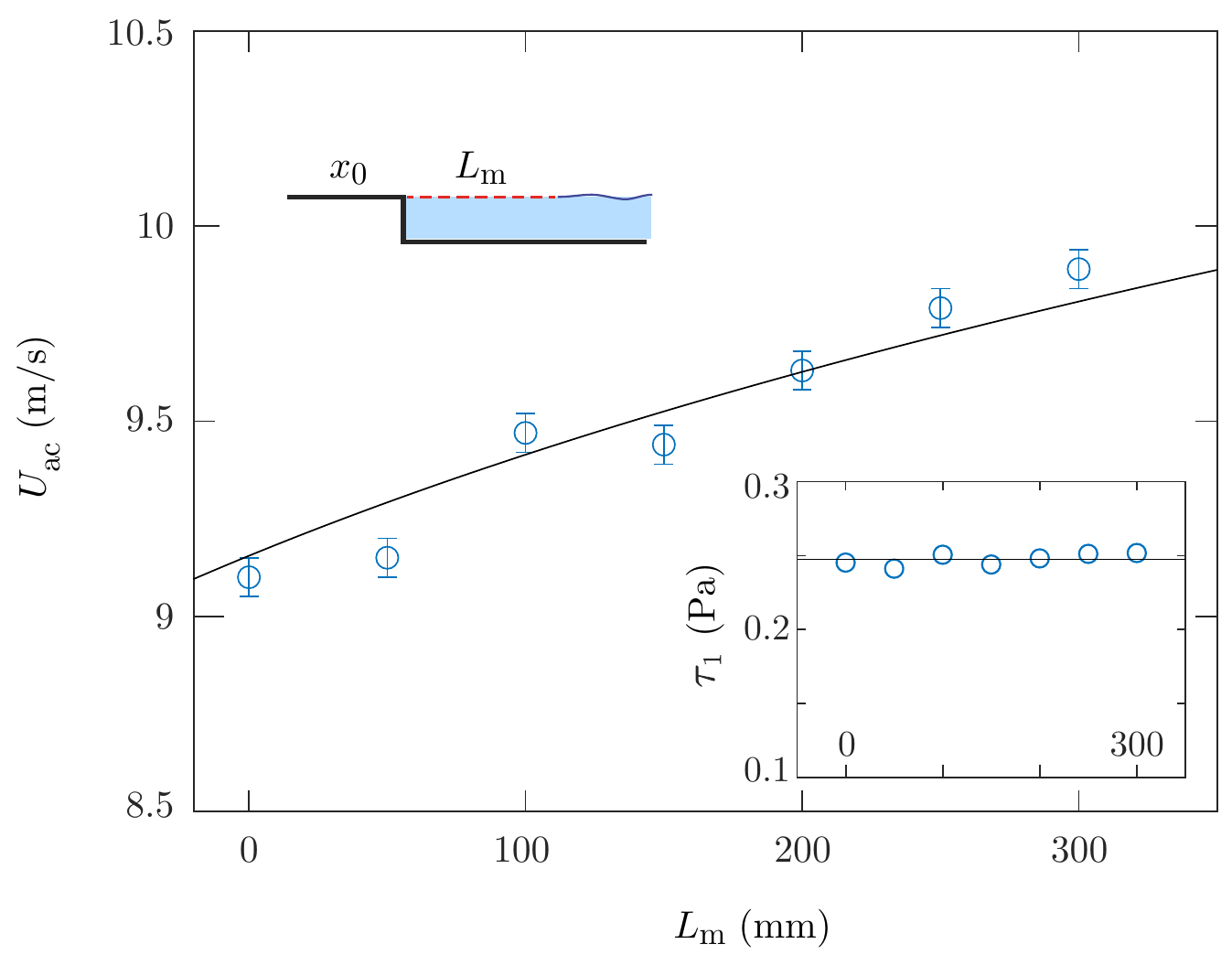}
\caption{Determination of the critical shear stress $\tau_1$ for the onset of solitons, obtained by measuring the critical wind velocity $U_{ac}$ as a function of the length $L_m$ of a floating rigid membrane that extends the length of the rigid-wall boundary layer (red dashed line in the sketch). From this measurement, the critical shear stress $\tau_1 \simeq 0.25$~Pa is obtained using the decay law (\ref{eq:tauvsx}), shown in the inset. The liquid depth  $h=35$~mm.}
\label{fig:tau1}
\end{center}
\end{figure}

The local shear stress $\tau(x)$ depends both on the wind velocity $U_a$ and on the distance $x$ along the channel through Eq.~(\ref{eq:tauvsx}). In order to determine the critical shear stress $\tau_1$ for the onset and instability of the initial wave packet, we varied the location of the initial wave packet by extending the  rigid wall upstream of the liquid surface using a rigid floating membrane of length $L_m$ fixed at the edge of the tank, at $x=0$ (see the sketch in Fig.~\ref{fig:tau1}). For each membrane length $L_m$, we measure the critical wind velocity $U_{ac}$, which typically increases from 9 to 10~\mps \ when the total length of the boundary layer is increased from $x_0=350$ to $x_0+L_m = 650$~mm. The critical shear stress deduced from this measurement using the decay law (\ref{eq:tauvsx}) is remarkably independent of the membrane length $L_m$ (see the inset of Fig.~\ref{fig:tau1}),
\begin{equation}
\tau_1 \simeq 0.25 \pm 0.01~\mathrm{Pa}.
\label{eq:tau1}
\end{equation}
This confirms that the shear stress, rather than the wind velocity, is the relevant control parameter for the onset of waves and solitons.

The secondary threshold $\tau_2$ can be estimated from the distance $x_2$ of ``natural decay'' of solitons, i.e. the location where isolated solitons spontaneously disappear in the absence of interaction with other solitons. This distance $x_2$ must be determined for a wind velocity close to the onset, when the emission frequency is small and solitons remain isolated [see Fig.~\ref{spatiotemp}(a)]. We find $x_{2}\simeq 500 \pm 50$~mm (this value is determined by visual inspection of solitons; LSP measurements are restricted to $x = 400$~mm), from which Eq.~(\ref{eq:tauvsx}) yields
\begin{equation}
\tau_2 \simeq 0.21 \pm 0.01~\mathrm{Pa}.
\label{eq:tau2}
\end{equation}
In spite of this narrow hysteresis window $[\tau_2, \tau_1]$ of 15\%, 
the shallow decrease of $\tau(x)$ allows for a extended subcritical region $[x_1, x_2]$ where solitons can propagate.

At larger wind velocity, although $x_2$ should increase in principle, solitons are found to decay before reaching this distance: Because of the larger emission frequency, more than one soliton is present in the tank at the same time and the wind perceived by a mature downstream soliton is reduced by a newly formed upstream soliton. Interactions between solitons are depicted by horizontal dashed arrows and circles in Fig.~\ref{spatiotemp}. At moderate wind velocity [Fig.~\ref{spatiotemp}(b)], a marked slowdown of the downstream soliton is observed each time a new soliton is emitted, illustrating the high sensitivity of solitons to changes in the local boundary layer induced by upstream solitons. Such a sheltering effect may even lead to the decay of the downstream soliton when the two solitons are close enough, typically when they are separated by less than 200-300~mm (100 times larger than the soliton amplitude).  At higher wind velocity [Fig.~\ref{spatiotemp}(c)], this sheltering effect is less pronounced: Only a weak slowdown is observed at moderate fetch, while solitons at larger fetch are essentially not affected and continue their course.


\section{Generation of solitons}

We now describe in more detail the instability of the initial wave packet that leads to the formation of viscous solitons.

The initial wave packet has a characteristic wavelength $\lambda \approx 18$ mm $\pm$ 2 mm and frequency $f \simeq 1.0 \pm 0.1$~Hz [see Figs.~\ref{spatiotemp}(d) and \ref{stacked}], with no significant dependence on liquid depth or wind velocity. This wavelength, which is about twice the capillary length $\lambda_c=2\pi\sqrt{\gamma/\rho g} \approx 9.4$ mm, is consistent with the one found by Francis over a wide range of liquid viscosity (see Ref.~\cite{Miles1959generation}). The phase velocity of these waves, $c \simeq 19 \pm 2$~mm~s$^{-1}$, is about one-tenth of the minimum phase velocity of inviscid gravity-capillary waves (one has $c_{\rm min} = 170$~mm~s$^{-1}$ for a fluid with identical density and surface tension but zero viscosity).

Importantly, the wavelength in the initial wave packet is smaller than the viscous cutoff $\lambda_0 = 2\pi \theta_0^{-2/3} (\nu_\ell^2/g)^{1/3} \simeq 24$~mm below which waves at a stress-free interface cannot propagate~\cite{Lamb, Leblond87}; here $\theta=\nu_\ell k^2/\sigma$ is the parameter which compares the dissipation timescale and the inviscid wave period $\sigma^{-1}$ of a wave number $k$, with $\theta>\theta_0 \simeq 1.31$ for overdamped waves (this viscous cutoff $\lambda_0$ is essentially independent of the liquid depth $h$ for the values considered here).  The waves in the initial wave packet are therefore overdamped waves, resulting from a balance between the forcing by the wind and the viscous dissipation. This is consistent with the fact that they do not cross the boundary $x_1$ beyond which the local shear stress is not sufficient to sustain them [see Fig.~\ref{fig:sketch}(a)].

\begin{figure}
	\begin{center}
\includegraphics[width=14cm]{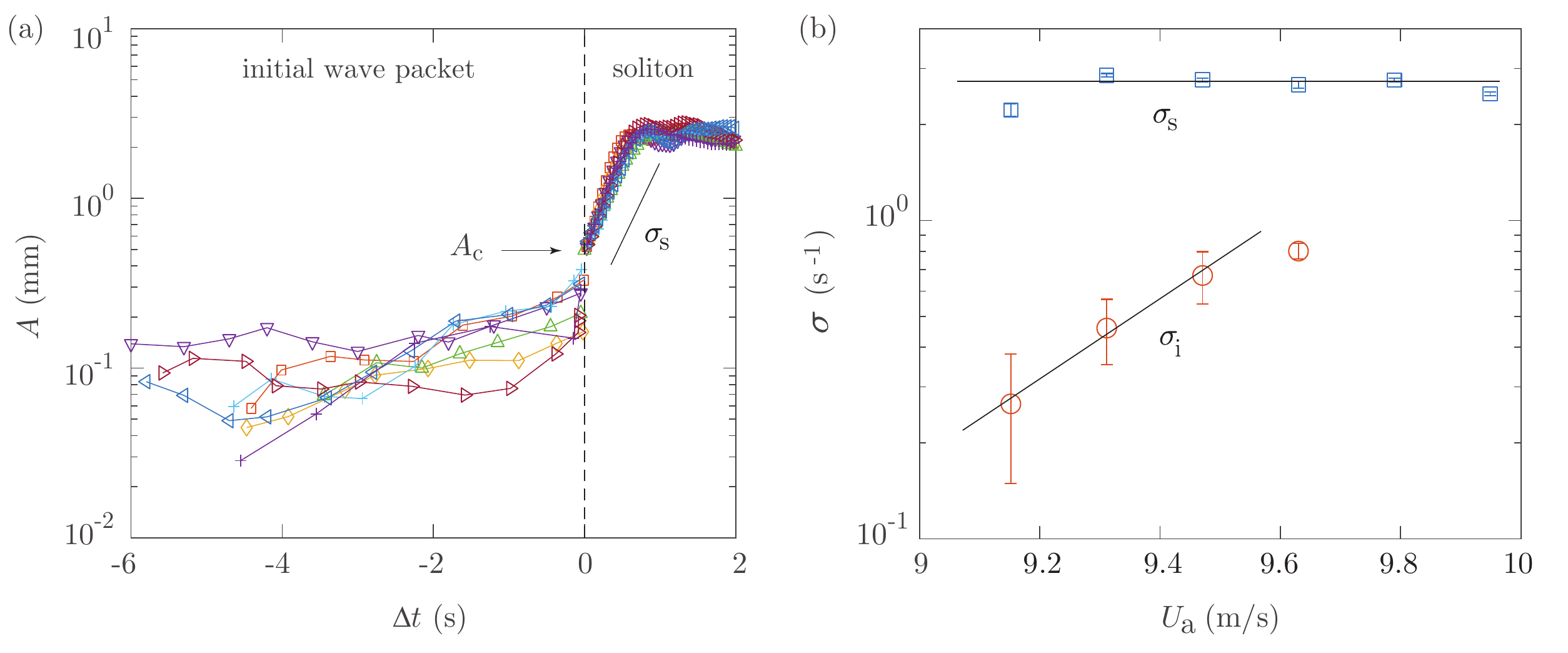}
\caption{(a) Amplitude of the initial wave packet and soliton crest as a function of the relative time $\Delta t = t - t_s$, for $U_a = 9.15$~\mps \ and $h=35$~mm. The time of formation of a soliton $t_s$ is defined such that $A = 0.5$~mm (see the arrow). For each event (different symbols), the data at $\Delta t<0$ shows the amplitude of the wave envelope at the center of the wave packet, $x_c=50$~mm, while the data at $\Delta t>0$ shows the amplitude following the soliton crest. The growth of the initial wave packet ($\Delta t<0$) is irregular, while the soliton growth ($\Delta t>0$) is reproducible, with a well-defined growth rate $\sigma_s$. (b)~Mean growth rate of the initial wave packet $\sigma_i$ just before the transition to the soliton and growth rate of the soliton $\sigma_s$ as functions of the wind velocity $U_a$.}
\label{fig:sigma_f_vs_ua}
\end{center}
\end{figure}

The initial wave packet, once it appears, is always unstable and forms a soliton. The formation of a soliton can be decomposed into two stages: a slow and erratic temporal growth of the initial wave packet, followed by a rapid steepening of its highest crest.  An instance of such erratic growth prior to the emission of a soliton is visible in the spatio-temporal diagram in Fig.~\ref{spatiotemp}(d) for the liquid depth $h=35$~mm --- see the slight temporal decrease of the wave amplitude at $t\simeq 31$~s. The variability of this first growth stage is more pronounced at lower liquid depth.

The time evolution of the amplitudes of the initial wave packet and soliton crest at the onset $U_a = U_{ac}$ is illustrated in Fig.~\ref{fig:sigma_f_vs_ua}(a). For each event, the data at $\Delta t = t-t_s<0$ shows  the amplitude of the wave envelope at a fixed point in the center of the wave packet, $x_c=50$~mm, while  the data at $\Delta t>0$ shows the amplitude following the soliton crest.  The transition time $t_s$ from wave to soliton is defined such that the wave envelope crosses a critical amplitude $A_c$, which we choose so as to minimize the dispersion in the soliton growth phase. We find $A_c \simeq 0.5 \pm 0.1$~mm, corresponding to a wave slope $kA_c \simeq 0.2 \pm 0.05$.

The rapid growth of the wave crest above a critical amplitude suggests that the transition is triggered by a modification in the air flow as the wave grows~\cite{belcher1993turbulent,belcher1998turbulent,sullivan2000simulation,Sullivan2018,Buckley2019}: For small amplitude the flow in the air closely follows the surface, resulting in a nearly symmetric pressure and shear stress distribution on both sides of the wave. As the amplitude increases, an asymmetry in the stress distribution gradually appears, resulting in an increased momentum transfer from the wind to the waves.

The growth rate $\sigma_s$ in the soliton phase, computed by fitting $A(t) \sim \exp(\sigma_s \, \Delta t)$ for $\Delta t >0$,  is almost independent of the wind velocity, $\sigma_s \simeq 2.8 \pm 0.3$~s$^{-1}$ [see Fig.~\ref{fig:sigma_f_vs_ua}(b)]. On the other hand, the growth of the initial wave packet prior to the formation of a soliton shows a high variability. We characterize this irregular growth by fitting $A(t) \sim \exp(\sigma_i \, \Delta t)$ for $\Delta t <0$ for each event right before the sharp increase at $A=A_c$ and we compute the average and standard deviation of $\sigma_i$ among the set of solitons. In spite of this large variability, we observe in Fig.~\ref{fig:sigma_f_vs_ua}(b) an increase of $\sigma_i$ with $U_a$. A similar increase is obtained for the two other liquid depths $h=20$ and 50~mm.

The nearly constant $\sigma_s$ and the increasing $\sigma_i$ with wind velocity are again consistent with a subcritical bifurcation: The initial growth is related to the linear instability of the base state, which is governed by the deviation from the critical shear stress $\tau-\tau_1$, whereas the dynamics when approaching the nonlinear soliton branch do not depend significantly on~$\tau$. The large variability in the initial wave packet evolution close to the onset probably originates from the temporal fluctuations of the shear stress. This is illustrated by the red trajectory in Fig.~\ref{fig:sketch}(b): For a mean shear stress close to the critical value $\tau_1$, temporal excursions of $\tau$ on both sides of $\tau_1$ are expected to produce an alternating positive and negative growth rate, and hence alternating amplification and attenuation of the wave amplitude. Such a cycle may repeat several times until a sufficiently long positive excursion leads to a jump on the bifurcated soliton branch.

\begin{figure}[t]
	\begin{center}
\includegraphics[width=14 cm]{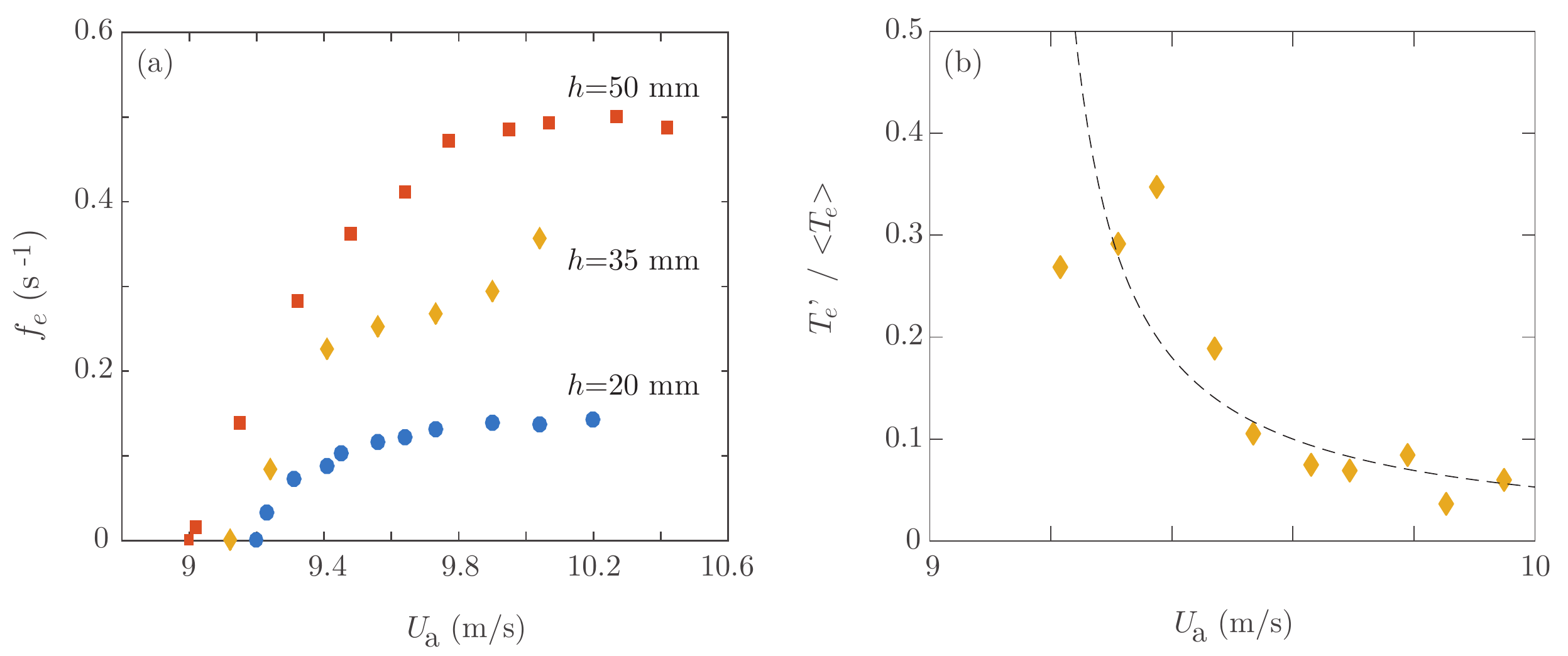}
\caption{(a) Emission frequency of solitons, for the three liquid depths. (b) Fluctuation rate of the waiting time $T_e$ between two solitons, for $h=35$~mm. The dashed line, $(U_a - U_{ac})^{-1}$ with $U_{ac} = 9.25$~\mps \ the critical wind velocity, is shown as a guide to the eye.}
\label{fe}
\end{center}
\end{figure}

After the emission of a soliton, the wave packet disappears for a certain time and then reforms. The waiting time $T_e$  between two solitons depends both on the time to ``refill'' the liquid tank after the emission of a soliton, and on the reformation and irregular growth time of the wave packet. Close to the onset, $T_e$ is expected to be governed by the second (slower) process, i.e. by the typical number of cycles performed around the critical shear stress $\tau_1$ before jumping to the soliton branch, which itself depends on the timescale of the fluctuating stress perceived by the wave compared to its growth time. The mean emission frequency $f_e = 1/\langle T_e\rangle$  and its fluctuation rate $T_e' / \langle T_e \rangle$, with $T_e'$ the standard deviation of $T_e$, are plotted in Fig.~\ref{fe} as a function of the wind velocity. The emission frequency starts from zero at the critical wind $U_a = U_{ac}$, with a large fluctuation rate,  and the system rapidly evolves towards a nearly periodic emission of solitons at a frequency $f_e$ at larger wind velocity. This nearly periodic emission far from the onset reflects the disappearance of the intermittent cycles when the instantaneous shear stress remains well above the threshold $\tau_1$ at all time.  The maximum emission frequency at large wind velocity is comparable to the wave frequency $f \simeq 1.0$~Hz. In the case $h=50$~mm, solitons are emitted at a particularly high rate $f_e \simeq f/2$: One out of two wave crests forms a soliton.

\section{Propagation and decay of mature solitons} 

\subsection{Amplitude and propagation velocity}

\begin{figure}[t]
\begin{center}
\includegraphics[width=14 cm]{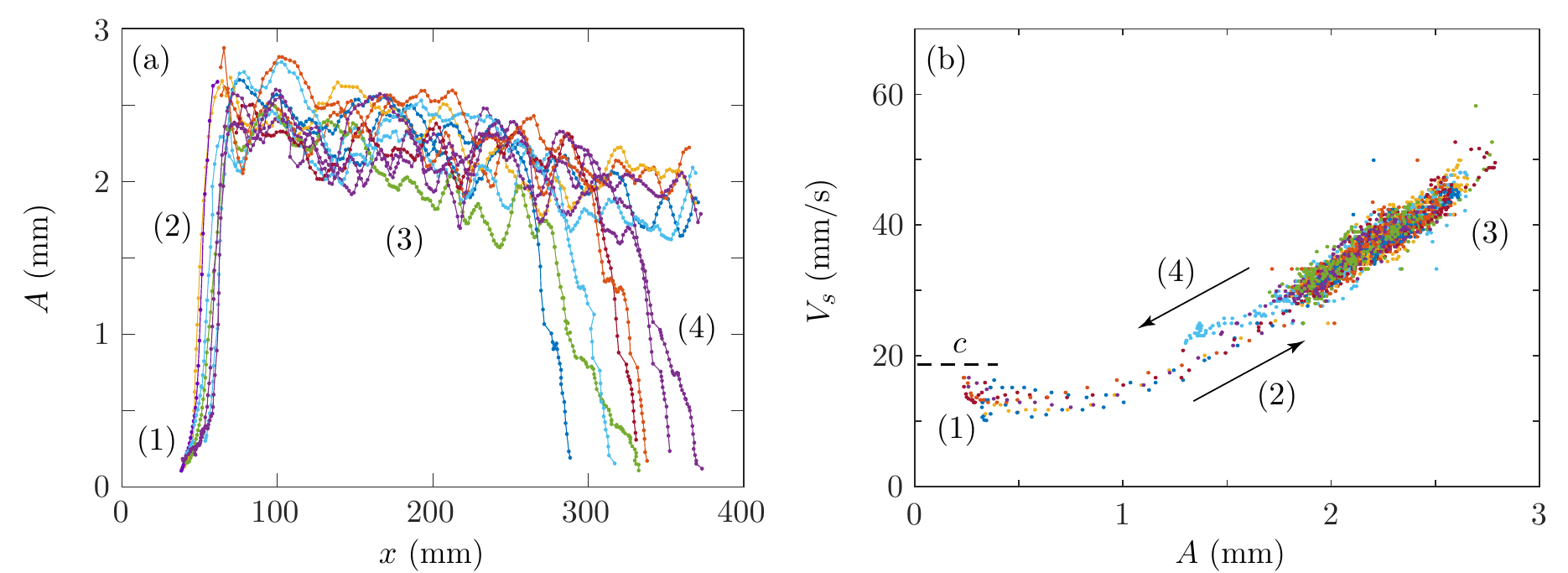}
\caption{(a) Spatial evolution of the soliton amplitude $A$ for a set of $N=12$ solitons, for a wind velocity $U_a=9.15$~m~s$^{-1}$ and liquid depth $h=35$ mm. (b) Soliton trajectories, in the plane amplitude $A$ vs velocity $V_s$. Each color represents a single soliton. (1) shows the unstable primary wave (2) the growth phase; (3) the slowly decrasing mature state, and (4) the rapid decay. The horizontal dashed line represents the mean phase velocity of the regular waves, $c \simeq 19$~mm~s$^{-1}$.}
 \label{a_vs_x}
\end{center}
\end{figure}

We now characterize the shape and propagation velocity of mature solitons in the subcritical region $x>x_1$.  We first plot in Fig.~\ref{a_vs_x}(a) the amplitude of a large set of solitons as a function of the distance $x$ along the channel. We recover the successive steps of their life cycle: (1) Solitons emerge from the initial wave packet at $x \simeq 50 \pm 10$~mm; (2) their amplitude strongly increases between 60 and 80~mm, where they reach their maximum value; (3)~during their ``mature'' state, their amplitude shows strong fluctuations ($\simeq \pm 10\%$) because of the stress fluctuations in the turbulent boundary layer, superimposed to a clear decreasing trend related to the decay of the mean shear stress along the channel; (4)~their amplitude rapidly falls at the end of their course. The fluctuations of amplitude in the mature phase (3) imply a large variability in the distance $x$ at which solitons disappear (note that the tracking of solitons is limited here to 380~mm, but a number of solitons continue their course up to $\simeq 500$~mm). The disappearance occurs on a short time scale of $1-2$~s, which is consistent with a sharp jump from the soliton branch to the base state when $\tau = \tau_2$ in the subcritical bifurcation diagram of Fig.~\ref{fig:sketch}(b). 

An interesting feature of viscous solitons is the strong correlation between their velocity $V_s$ and their amplitude $A$. This is shown in Fig.~\ref{a_vs_x}(b), where we plot the time history of a set of solitons in the plane $(A, V_s)$. Each color denotes a single soliton, sampled at 20~Hz. In phase (1), the soliton precursor (dominant wave crest of the regular wave packet) has a velocity independent of its amplitude, $V_s \simeq 13 \pm 3$~mm~s$^{-1}$, a value slightly smaller than the average phase velocity of regular waves, $c \simeq 19 \pm 2$~mm~s$^{-1}$. However, as the amplitude increases above 1~mm, $A$ and $V_s$ increase simultaneously, reach their maximum mature values, and then slowly decrease. In the decay phase (4), $A$ and $V_s$ rapidly fall, approximately following path (2) in the opposite direction.

\begin{figure}
	\begin{center}
\includegraphics[width=14 cm]{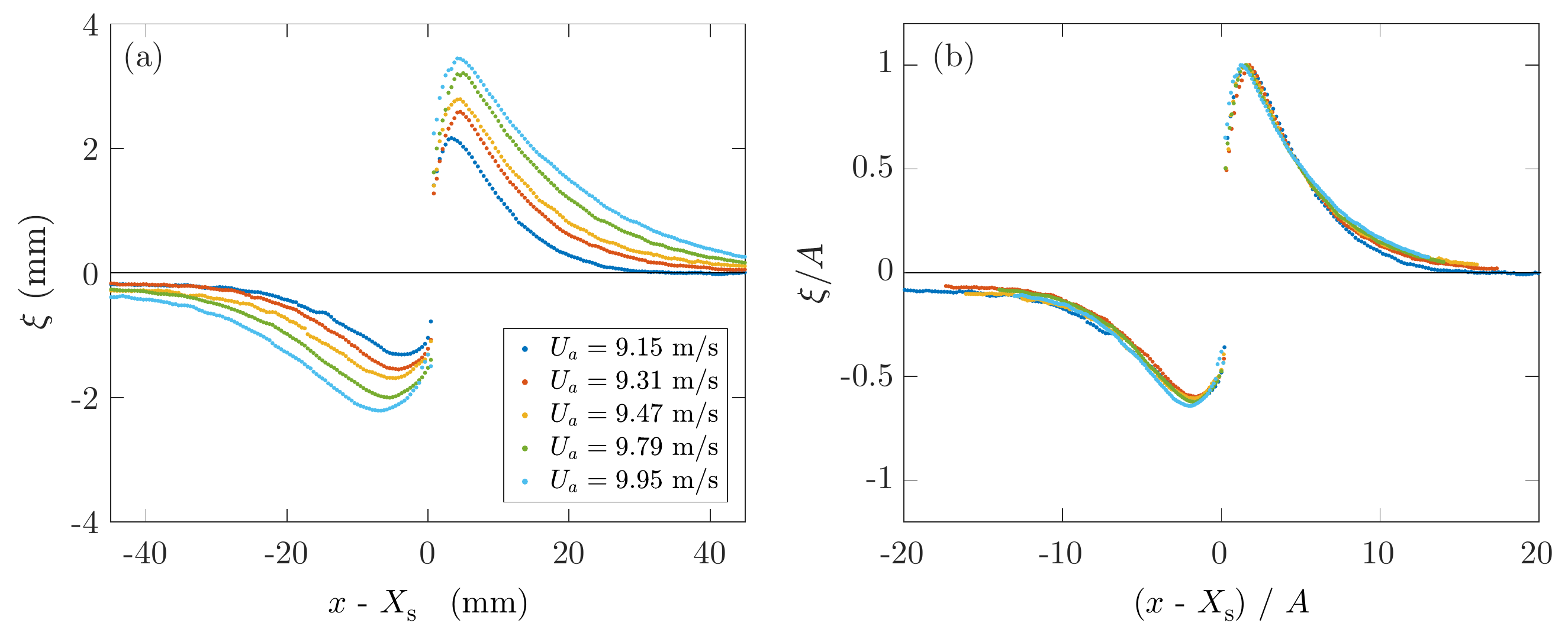}
\caption{Shape of mature solitons at different wind velocities $U_a$, for a liquid depth $h=35$~mm. Each curve is an average over a set of solitons at constant $U_a$, with the center $X_s$ defined such that $\xi(X_s) = 0$. (a) Profiles $\xi$ in physical units. (b) Profiles rescaled by the soliton amplitude $A = \max(\xi)$. Note that the measurements at $x \simeq X_s$ are not reliable, where the true slope may become negative (overhanging profile).}
\label{forme_soliton}
\end{center}
\end{figure}

During the mature phase, although the amplitude of the solitons fluctuates and slightly decreases, their shape remains unchanged. The average shape of mature solitons is shown in Fig.~\ref{forme_soliton}(a) for various wind velocities. Each curve represents an average of soliton profiles $\xi$ shifted about their center $X_s$ in their early mature phase (averaged over an interval of time corresponding to their first 100~mm of propagation). When both the height $\xi$ and spatial coordinate $x$ are rescaled with the maximum amplitude $A = \max (\xi)$, the profiles collapse onto a single master curve [Fig.~\ref{forme_soliton}(b)]: The soliton shape is remarkably self-similar, at least not too close to the center $X_s \simeq 0$ where measurements are not reliable. The crest is at $\Delta x \simeq 1.5 A$ and the trough ($\min \xi \simeq -0.6 A$) is at $\Delta x \simeq -2A$, yielding a trough-to-crest mean slope of order of 0.5. The local slope at the center is in reality much larger, and may even become negative. We also note that although the crest is larger than the trough, the front side decays more rapidly than the rear side, so the integral of the surface elevation over the whole soliton profile is essentially zero: The mass transported by the soliton crest is compensated by the long negative tail, which corresponds to the slow upstream refilling by the recirculation flow.

\begin{figure}
	\begin{center}
\includegraphics[width=13cm]{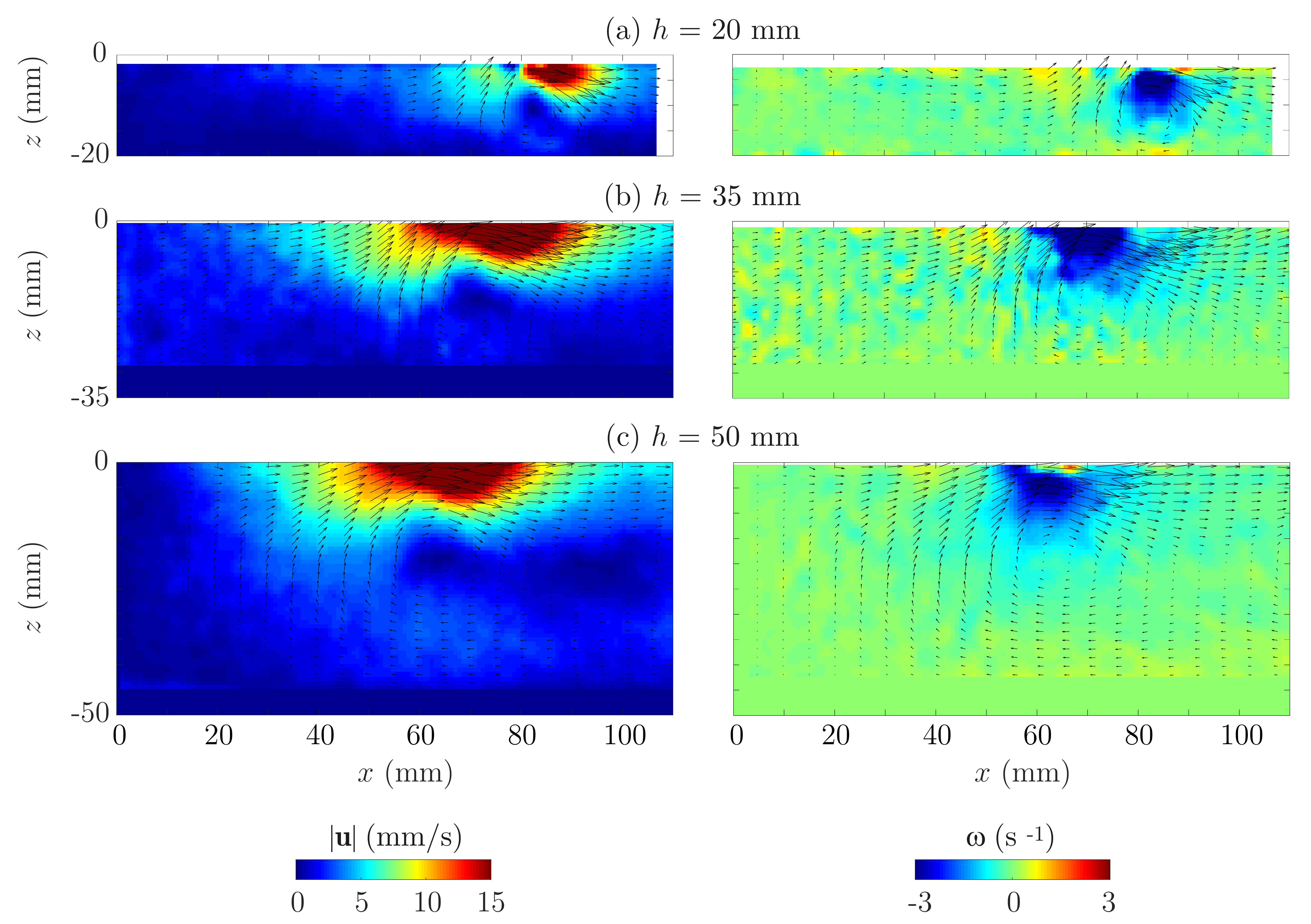}
\caption{Flow below a soliton for different liquid depths (a) $h=20$~mm, (b) $h=35$~mm, and (c) $h=50$~mm (wind velocity $U_a = 10.1$~\mps). The color maps show the norm of the velocity $|{\bf u}| = (u_x^2 + u_z^2)^{1/2}$ (left column) and the vorticity $\omega$ (right column). The selected fields are shown just after the emission, at the beginning of the mature phase (note that for small $h$ the emission occurs at a larger fetch).}
\label{fig:PIVvsh}
\end{center}
\end{figure}

\begin{figure}
	\begin{center}
\includegraphics[width=14cm]{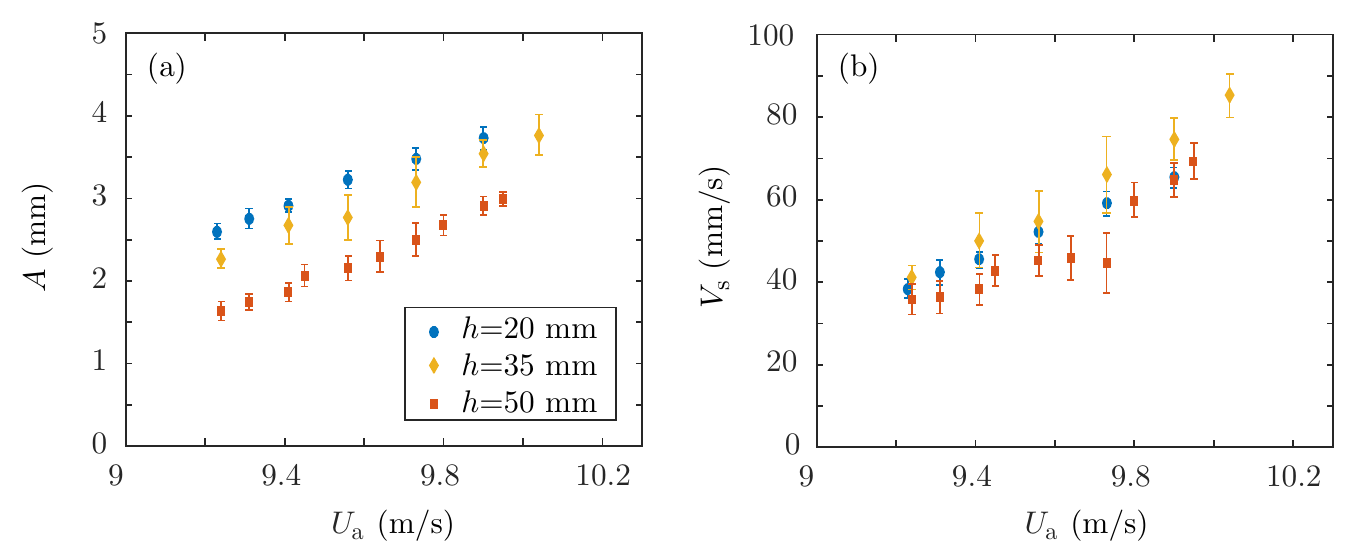}
\caption{(a) Mean velocity $V_s$  and (b) mean amplitude $A$ of solitons in the quasistationary phase as a function of the wind velocity for various liquid depths $h$.}
\label{A+deUa}
\end{center}
\end{figure}

The structure of the flow below solitons is shown in Fig.~\ref{fig:PIVvsh} for three liquid depths.  The vortex is located at $-z/h \simeq 0.4-0.5$, which is slightly deeper than the depth $-z/h=1/3$ at which the velocity of the mean recirculation flow vanishes [see Eq.~(\ref{eq:par})]. The vorticity is concentrated in a layer of thickness of order 1~cm below the surface. Surprisingly, decreasing the liquid depth $h$ tends to produce solitons of larger amplitude, but has a weak influence on their propagation velocity $V_s$ (Fig.~\ref{A+deUa}). This dependence on $h$ suggests that the bottom of the tank plays an important role in the formation of solitons. The decrease of the amplitude for increasing depth raises the question of the existence of solitons for larger depth, but it cannot be tested in the present setup. Producing even higher solitons by decreasing the liquid depth below 20~mm is not possible either,  because the mass transported by the crest of solitons cannot be compensated by the too slow return flow, leading to a gradual flushing of the tank.

\subsection{Force balance on mature solitons}

We show now that, in the quasi-stationary mature phase, solitons are nonlinear objects pushed by the wind: Their propagation velocity $V_s$ is proportional to their amplitude $A$, as the result of a balance between the inertial drag on the air side and the viscous drag on the liquid side.

On the air side, the aerodynamic drag is turbulent: The soliton height, $2-4$~mm, is 100 times larger than the thickness of the viscous sublayer ($\delta_\nu = \nu_a / u^* \simeq 30~\mu$m at this wind velocity, with $u^* = \sqrt{\tau / \rho_a}$ the friction velocity). The wind velocity at the top of a soliton is therefore essentially given by $U_a$, with a typical Reynolds number of $U_a A/\nu_a \simeq 2000$. The drag force exerted by the wind on the soliton is therefore of the form
\begin{equation}
F_D \simeq C_D\rho_a U_a^2 W A,
\label{eq:fd}
\end{equation}
with $W$ the lateral extent of the soliton and $C_D$ a drag coefficient that depends only on its shape.

On the liquid side, the flow is laminar (the maximum Reynolds number is $V_s A / \nu_\ell \simeq 0.3$). We can therefore model the viscous friction by a Stokes drag force $F_\nu \simeq S \tau_s$, with $S \simeq A W$ the surface area of the soliton and $\tau_s$ the viscous stress. Since the velocity gradients remain concentrated in a layer of thickness of order $A$ below the surface (see Fig.~\ref{fig:PIVvsh}), we can write $\tau_s \simeq \rho_\ell \nu_\ell V_s / A$, yielding
\begin{equation}
F_\nu \simeq \beta \rho_\ell \nu_\ell W V_s,
\label{eq:fnu}
\end{equation}
where $\beta$ is a numerical factor of order 1, which may depend on the shape of the soliton and the liquid depth.

\begin{figure}
	\begin{center}
\includegraphics[width=8cm]{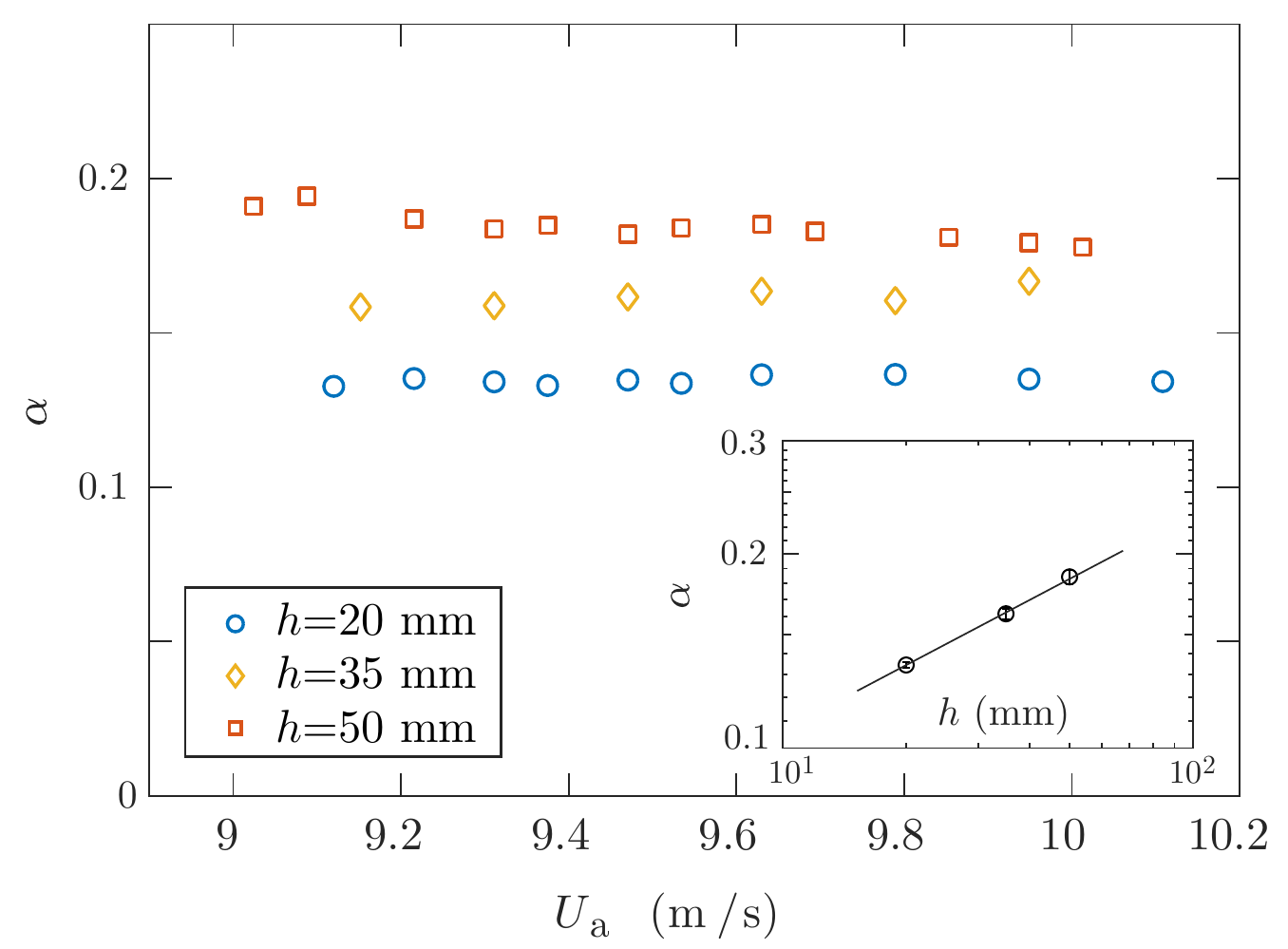}
\caption{Coefficient $\alpha$ characterizing the soliton velocity in Eq.~(\ref{eq:alpha}) as a function of the wind velocity $U_a$ for the three liquid depths $h$. The inset shows the mean value of $\alpha$ as a function of $h$, with a best fit $\alpha \sim h^{0.3}$.}
\label{alpha}
\end{center}
\end{figure}

Balancing the inertial force (\ref{eq:fd}) and the viscous force (\ref{eq:fnu}) finally selects the velocity of the soliton in the mature phase,
\begin{equation}
V_s \simeq \alpha \frac{\rho_a}{\rho_\ell} \frac{U_a^2 A}{\nu_\ell},
\label{eq:alpha}
\end{equation}
where $\alpha=C_D/\beta$ is a numerical factor of order 1.  This scaling is confirmed by the measured values of the coefficient $\alpha$ shown in  Fig.~\ref{alpha}, which is independent of $U_a$ for the whole range of velocity where solitons are observed.  The mean value of $\alpha$ slightly increases with the liquid depth ($\alpha \sim h^{0.3}$, see the inset): This trend originates from an increase of the viscous drag as $h$ is decreased, resulting in an increased drag coefficient $\beta$ in Eq.~(\ref{eq:fnu}).  Note that linearity between $A$ and $V_s$ applies only in the quasi-stationary mature phase (3), not in the growth (2) and decay (4) phases [see Fig.~\ref{a_vs_x}(b)].

\subsection{Reconstruction of the soliton branch}

\begin{figure}
\begin{center}
\includegraphics[width=8cm]{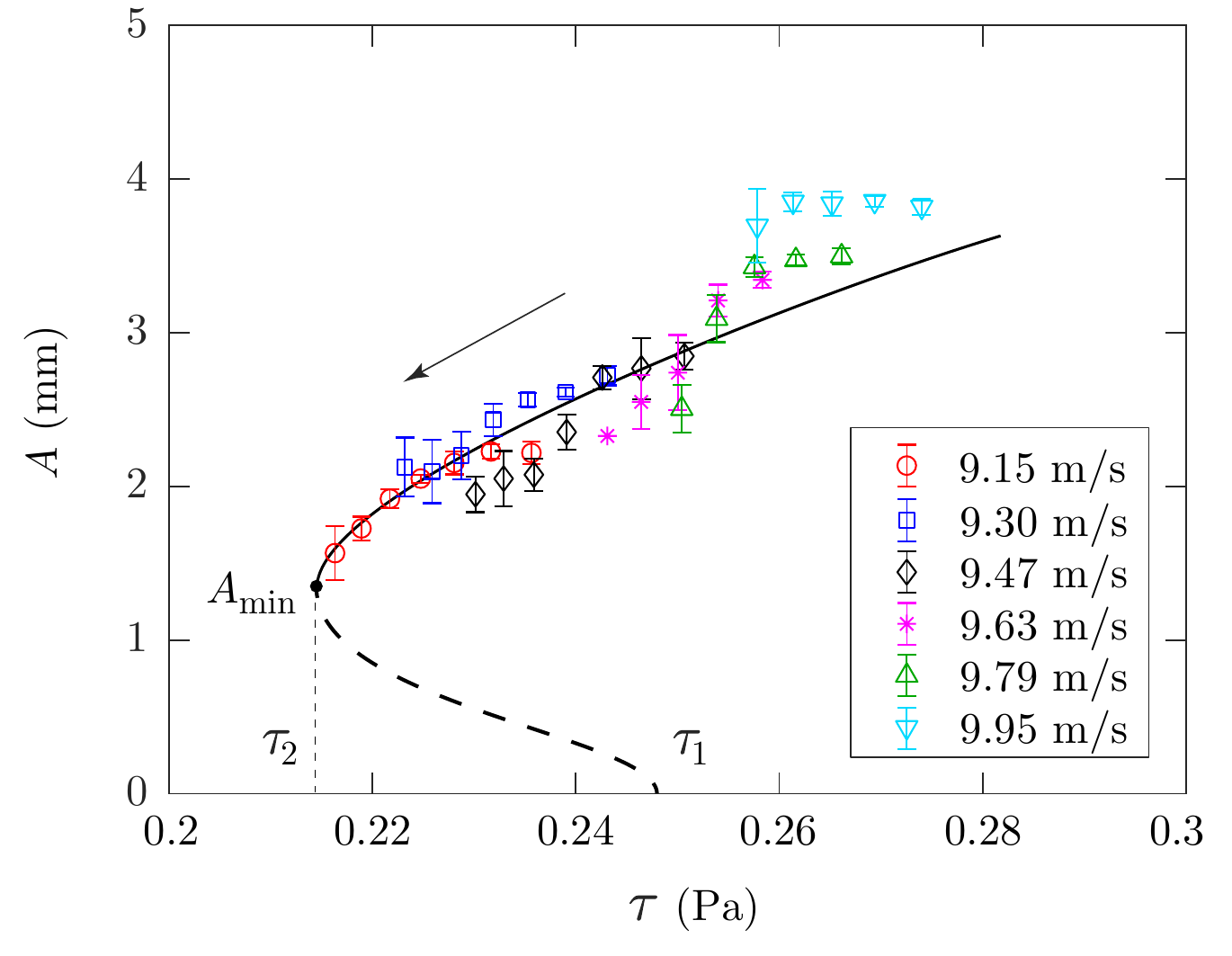}
\caption{Amplitude of viscous solitons as a function of the local shear stress $\tau$ for various wind velocities, obtained from the distance $x$ along the channel using the decay law (\ref{eq:tauvsx}). The distance $x$ increases from right to left in this diagram. The critical shear stresses $\tau_1$ and $\tau_2$ are from Eqs.~(\ref{eq:tau1}) and (\ref{eq:tau2}). The dashed line is a guide to the eye representing the hypothetical unstable branch of the subcritical bifurcation diagram of Fig.~\ref{fig:sketch}(b).}
\label{fig:Avstau}
\end{center}
\end{figure}

We finally show that from the slow decay of the soliton amplitude in the mature phase [Fig.~\ref{a_vs_x}(a)], it is possible to reconstruct the stable soliton branch in the subcritical bifurcation diagram hypothesized in Fig.~\ref{fig:sketch}(b).  Such bifurcation diagram implicitly assumes that the local shear stress uniquely selects the properties of the solitons, i.e. it ignores the retroaction of the soliton on the shear stress. To test to what extent this assumption holds, we plot in Fig.~\ref{fig:Avstau} the mean soliton amplitude as a function of the local shear stress $\tau$, deduced from the wind velocity and the location $x$ along the tank using the decay law (\ref{eq:tauvsx}). The data collapses reasonably well onto a single curve, which confirms that the soliton amplitude is correctly described by the local shear stress. In this plot, the unstable branch between $\tau_1$ and $\tau_2$ (shown as a dashed line) is drawn as a guide to the eye, but it cannot be accessed in this experiment.

From Fig.~\ref{fig:Avstau} we can summarize the evolution of a soliton as follows: For a given wind velocity, provided the local shear stress is above $\tau_1$, the system first jumps to the soliton branch $A \neq 0$ (solid line), follows the branch from right to left for increasing $x$, and finally jumps back to the stable branch $A=0$ when the local shear stress reaches $\tau_2$.
This return to the base state is either natural, when the soliton reaches the distance $x_2$ where $\tau(x)=\tau_2$, or forced, when the air drag on the soliton is reduced by the emission of a new upstream soliton.

\section{Conclusion}
\label{sec:conclusion}

In this paper we characterized experimentally the viscous solitons first observed by Francis~\cite{Francis_1954}, that are formed when wind blows over the surface of a highly viscous liquid.  These out-of-equilibrium structures result from a balance between the forcing by the wind and the viscous dissipation in the liquid, which selects a non-trivial self-preserving asymmetric shape. Our results suggest that viscous solitons arise from an initial unstable wave train when the wave amplitude  becomes large, triggering a change in the air flow that abruptly increases the aerodynamic drag on the wave.

We proposed here a simple model to describe the growth, propagation and decay of viscous solitons, based on a subcritical bifurcation governed by the local shear stress. A key ingredient to capture their life cycle is the streamwise decay of the local shear stress applied on the liquid surface, due to the development of the turbulent boundary layer in the air. Although this particular inhomogeneity of the shear stress field is specific to our flow configuration, we believe that viscous solitons are generic objects that could be observed in other configurations. Using a fully developed channel flow instead of a developing boundary layer would provide an interesting configuration with a uniform shear stress. For an applied shear stress $\tau$ inside the hysteresis window $[\tau_2, \tau_1]$, a single soliton propagating to infinity could be in principle generated by means of a localized mechanical forcing. Such a fully developed configuration with uniform shear stress is encountered in two-phase pipe flows, but the strong confinement usually present in this configuration (soliton amplitude of the order of the gas layer thickness) rapidly leads to the formation of slugs rather than solitons~\cite{andritsos1989effect}.

Viscous solitons resulting from a balance between wind forcing and viscous dissipation are also found in the Hele-Shaw geometry~\cite{Gondret97b, Meignin2013}. In this configuration, the two-phase flow confined in the thin gap between two vertical parallel plates generates propagating localized structures resulting from a subcritical instability. Because of the transverse confinement, the flow in a Hele-Shaw cell is laminar in both phases, and the small gap size implies a thin uniform shear layer between the fluids, thus mimicking a quasi-inviscid discontinuous velocity profile. The nature of the force balance in the quasi-two-dimensional (quasi-2D) Hele-Shaw flow strongly differs  from the 3D configuration: The capillary and the friction forces are governed by the transverse scale (gap size) in the quasi-2D case, rather than by the vertical scale (soliton amplitude) in the 3D case. In spite of these differences, solitons with similar asymmetric shapes are found in both configurations,  with an almost vertical rear facing the wind and a weak slope at the front.

The subcritical nature of the instability that leads to the generation of viscous solitons raises a number of questions. How does the hysteresis window $[\tau_2, \tau_1]$, which is about 15\% in the present experiments, depend on the liquid properties and flow parameters? Under what conditions is this transition subcritical? More precisely, for a given flow configuration, is there a critical liquid viscosity $\nu_{\ell c}$ above which the instability becomes subcritical? The experiments of Paquier~{\it et al.} \cite{Paquier_2016} suggest a transition for $\nu_{\ell c} \simeq 100-200$~mm$^2$~s$^{-1}$, but the dependence of this critical viscosity on the other parameters remains unexplored. If only the liquid properties are considered, i.e., ignoring the possible influence of the liquid depth $h$, dimensional analysis implies
\begin{equation}
\nu_{\ell c} \simeq \left( \frac{\gamma^3}{\rho_\ell^3 g} \right)^{1/4}.
\label{eq:nuc}
\end{equation}
We obtain $\nu_{\ell c} \simeq 180$~mm$^2$~s$^{-1}$ for silicon oil, which turns out to provide the correct order of magnitude for the onset of solitons. We note that for this viscosity, the viscous cutoff for linear waves $(\nu_\ell^2/g)^{1/3}$ matches the capillary length $(\gamma/\rho_\ell g)^{1/2}$, suggesting that the instability becomes subcritical and generates solitons when the most unstable wavelength is critically damped and cannot propagate.

We conclude this paper by conjecturing that a necessary condition to form viscous solitons by blowing over a viscous liquid is to first generate critically damped waves of sufficient slope, over which flow separation occurs. For waves of wavelength of order of the capillary length, such as those naturally arising from the Kelvin-Helmholtz instability, this condition requires a liquid of viscosity larger than $\nu_{\ell c}$ [Eq.~(\ref{eq:nuc})]. This conjecture is however difficult to check experimentally because this critical viscosity  cannot be varied significantly using ordinary liquids.

We finally note a striking similarity between viscous solitons and the hydroelastic localized structures that form at the surface of a compliant gel forced by wind \cite{hansen1980,gad1984interaction, kim2014space}. The turbulent air flow above a wavy viscous liquid and a wavy soft solid being similar, this suggests that the physical mechanism responsible for the generation of solitons (detached flow over overdamped waves) is not specific to viscous liquids, but is also present in viscoelastic solids.

\begin{acknowledgments}

We are grateful to F. Charru, G. Dietze, P. Gondret, A. Paquier, and S. Perrard for fruitful discussions. We thank  A. Limare for help in selecting pigments used in laser sheet profilometry, and A. Aubertin, L. Auffray, J. Amarni and R. Pidoux for experimental help.
This work was supported by the project ``ViscousWindWaves'' (Project No. ANR-18-CE30-0003) of the French National Research Agency.

\end{acknowledgments}

\bibliographystyle{unsrt}
\bibliography{Biblio_Solitons}

\begin{thebibliography}{10}

\bibitem{Kahma_1988}
K.~Kahma and M.~A. Donelan.
\newblock A laboratory study of the minimum wind speed for wind wave
  generation.
\newblock {\em J. Fluid Mech.}, 192:339--364, 1988.

\bibitem{caulliez1999three}
G.~Caulliez and F.~Collard.
\newblock Three-dimensional evolution of wind waves from gravity-capillary to
  short gravity range.
\newblock {\em European Journal of Mechanics-B/Fluids}, 18(3):389--402, 1999.

\bibitem{lin2008direct}
M.-Y. Lin, C.-H. Moeng, W.-T. Tsai, P.~P. Sullivan, and S.~E. Belcher.
\newblock Direct numerical simulation of wind-wave generation processes.
\newblock {\em J. Fluid Mech.}, 616:1--30, 2008.

\bibitem{Paquier_2015}
A.~Paquier, F.~Moisy, and M.~Rabaud.
\newblock Surface deformations and wave generation by wind blowing over a
  viscous liquid.
\newblock {\em Phys. Fluids}, 27:122103, 2015.

\bibitem{zavadsky2017two}
A.~Zavadsky, A.~Benetazzo, and L.~Shemer.
\newblock On the two-dimensional structure of short gravity waves in a wind
  wave tank.
\newblock {\em Physics of Fluids}, 29(1):016601, 2017.

\bibitem{Perrard2019}
S.~Perrard, A.~Lozano-Dur{\'a}n, M.~Rabaud, M.~Benzaquen, and F.~Moisy.
\newblock Turbulent windprint on a liquid surface.
\newblock {\em Journal of Fluid Mechanics}, 873:1020--1054, 2019.

\bibitem{hwang2019wind}
P.~A. Hwang, D.~W. Wang, J.~Yungel, R.~N. Swift, and W.~B. Krabill.
\newblock Do wind-generated waves under steady forcing propagate primarily in
  the downwind direction?
\newblock {\em arXiv preprint arXiv:1907.01532}, 2019.

\bibitem{dias1999nonlinear}
F.~Dias and C.~Kharif.
\newblock Nonlinear gravity and capillary-gravity waves.
\newblock {\em Annual review of fluid mechanics}, 31(1):301--346, 1999.

\bibitem{veron2001experiments}
F.~Veron and W.~K. Melville.
\newblock Experiments on the stability and transition of wind-driven water
  surfaces.
\newblock {\em J. Fluid Mech.}, 446(10):25--65, 2001.

\bibitem{zavadsky2017investigation}
A~Zavadsky and L~Shemer.
\newblock Investigation of statistical parameters of the evolving wind wave
  field using a laser slope gauge.
\newblock {\em Physics of Fluids}, 29(5):056602, 2017.

\bibitem{Paquier_2016}
A.~Paquier, F.~Moisy, and M.~Rabaud.
\newblock Viscosity effects in wind wave generation.
\newblock {\em Phys. Rev. Fluids}, 1:083901, 2016.

\bibitem{christov1995dissipative}
C.~I. Christov and M.~G. Velarde.
\newblock Dissipative solitons.
\newblock {\em Physica D: Nonlinear Phenomena}, 86(1-2):323--347, 1995.

\bibitem{Knobloch2008}
E.~Knobloch.
\newblock Spatially localized structures in dissipative systems: open problems.
\newblock {\em Nonlinearity}, 21:T45--T60, 2008.

\bibitem{Knobloch2015}
E.~Knobloch.
\newblock Spatial localization in dissipative systems.
\newblock {\em Annual Review of Condensed Matter Physics}, 6:325--359, 2015.

\bibitem{Francis_1954}
J.~R.~D. Francis.
\newblock Wave motions and the aerodynamic drag on a free oil surface.
\newblock {\em Phil. Mag.}, 45:695--702, 1954.

\bibitem{francis1956lxix}
J.~R.~D. Francis.
\newblock {LXIX}. {C}orrespondence. {W}ave motions on a free oil surface.
\newblock {\em Philosophical Magazine}, 1(7):685--688, 1956.

\bibitem{Miles1959generation}
J.~W. Miles.
\newblock On the generation of surface waves by shear flows. {P}art 3.
  {K}elvin-{H}elmholtz instability.
\newblock {\em J. Fluid Mech.}, 6(04):583--598, 1959.

\bibitem{chandrasekhar}
S.~Chandrasekhar.
\newblock {\em Hydrodynamic and hydromagnetic stability}.
\newblock Dover, New-York, 1961.

\bibitem{Paquier_PhD_2016}
A.~Paquier.
\newblock {\em Generation and growth of wind waves over a viscous liquid}.
\newblock PhD thesis, University Paris-Saclay, July 2016.

\bibitem{boomkamp1996classification}
P.~Boomkamp and R.~Miesen.
\newblock Classification of instabilities in parallel two-phase flow.
\newblock {\em International Journal of Multiphase Flow}, 22:67--88, 1996.

\bibitem{taylor1940generation}
G.~I. Taylor.
\newblock Generation of ripples by wind blowing over a viscous fluid.
\newblock {\em The Scientific Papers of G. I. Taylor}, 3:244--254, 1940.

\bibitem{Hogan1985}
J.~M. Hogan and P.~S. Ayyaswamy.
\newblock Linear stability of a viscous--inviscid interface.
\newblock {\em Physics of Fluids}, 28:2709, 1985.

\bibitem{barnea1993kelvin}
D.~Barnea and Y.~Taitel.
\newblock Kelvin-{H}elmholtz stability criteria for stratified flow: viscous
  versus non-viscous (inviscid) approaches.
\newblock {\em International journal of multiphase flow}, 19(4):639--649, 1993.

\bibitem{kim2011viscous}
H.~Kim, J.~C. Padrino, and D.~D. Joseph.
\newblock Viscous effects on {K}elvin--{H}elmholtz instability in a channel.
\newblock {\em J. Fluid Mech.}, 680:398--416, 2011.

\bibitem{andritsos1989effect}
N.~Andritsos, L.~Williams, and T.~J. Hanratty.
\newblock Effect of liquid viscosity on the stratified-slug transition in
  horizontal pipe flow.
\newblock {\em International Journal of Multiphase Flow}, 15(6):877--892, 1989.

\bibitem{Schlichtling}
H.~Schlichtling.
\newblock {\em Boundary Layer Theory}.
\newblock Springer, 8th edition, 2000.

\bibitem{Lamb}
Sir~H. Lamb.
\newblock {\em Hydrodynamics}.
\newblock Sixth edition, Cambridge University Press, 1995.

\bibitem{Leblond87}
P.~H. LeBlond and F.~Mainardi.
\newblock The viscous damping of capillary-gravity waves.
\newblock {\em Acta Mechanica}, 68:203--222, 1987.

\bibitem{belcher1993turbulent}
S.E. Belcher and J.C.R. Hunt.
\newblock Turbulent shear flow over slowly moving waves.
\newblock {\em Journal of Fluid Mechanics}, 251:109--148, 1993.

\bibitem{belcher1998turbulent}
S.E. Belcher and J.C.R. Hunt.
\newblock Turbulent flow over hills and waves.
\newblock {\em Annual Review of Fluid Mechanics}, 30(1):507--538, 1998.

\bibitem{sullivan2000simulation}
P.~P. Sullivan, J.C. McWilliams, and C.H. Moeng.
\newblock Simulation of turbulent flow over idealized water waves.
\newblock {\em Journal of Fluid Mechanics}, 404:47--85, 2000.

\bibitem{Sullivan2018}
P.P. Sullivan, M.L. Banner, R.P. Morison, and W.L. Peirson.
\newblock Turbulent flow over steep steady and unsteady waves under strong wind
  forcing.
\newblock {\em Journal of Physical Oceanography}, 48:3--27, 2018.

\bibitem{Buckley2019}
M.~P. Buckley and F.~Veron.
\newblock The turbulent airflow over wind generated surface waves.
\newblock {\em European Journal of Mechanics - B/Fluids}, 73:132--143, 2019.

\bibitem{Gondret97b}
P.~Gondret and M.~Rabaud.
\newblock Shear instability of two-fluid parallel flow in a {H}ele-{S}haw cell.
\newblock {\em Phys. Fluids}, 9:3267--3274, 1997.

\bibitem{Meignin2013}
L.~Meignin, P.~Gondret, C.~Ruyer-Quil, and M.~Rabaud.
\newblock Subcritical {K}elvin-{H}elmholtz instability in a {H}ele-{S}haw cell.
\newblock {\em Phys. Rev. Letters}, 90:234502, 2003.

\bibitem{hansen1980}
R.~Hansen, D.~Hunston, C.~Ni, and M.~Reischman.
\newblock An experimental study of flow-generated waves on a flexible surface.
\newblock {\em Journal of Sound and Vibration}, 68(3):317---334, 1980.

\bibitem{gad1984interaction}
M.~Gad-El-Hak, R.~F. Blackwelder, and J.~J. Riley.
\newblock On the interaction of compliant coatings with boundary-layer flows.
\newblock {\em J. Fluid Mech.}, 140:257--280, 1984.

\bibitem{kim2014space}
E.~Kim and H.~Choi.
\newblock Space-time characteristics of a compliant wall in a turbulent channel
  flow.
\newblock {\em J. Fluid Mech.}, 756:30, 2014.

\end{thebibliography}

\end{document}